\documentclass[a4paper,12pt]{article}
\usepackage[sumlimits,intlimits,namelimits]{amsmath}
\usepackage{amssymb}
\usepackage{a4wide}
\usepackage{comment}
\usepackage{xcolor}
\usepackage{multirow}
\usepackage{nicefrac}
\usepackage{slashed}

\usepackage[english]{babel}
\makeatletter
\def\fmslash{\@ifnextchar[{\fmsl@sh}{\fmsl@sh[0mu]}}
\def\fmsl@sh[#1]#2{%
  \mathchoice
    {\@fmsl@sh\displaystyle{#1}{#2}}%
    {\@fmsl@sh\textstyle{#1}{#2}}%
    {\@fmsl@sh\scriptstyle{#1}{#2}}%
    {\@fmsl@sh\scriptscriptstyle{#1}{#2}}}
\def\@fmsl@sh#1#2#3{\m@th\ooalign{$\hfil#1\mkern#2/\hfil$\crcr$#1#3$}}
\makeatother
\definecolor{asparagus}{rgb}{0.53, 0.66, 0.42}
\usepackage{graphicx}
\usepackage{subfigure}
\usepackage{epstopdf}
\newcommand{\muGperp}{{\mu_G^\perp}}
\newcommand{\muPperp}{{\mu_\pi^\perp}}

\begin{document}
\begin{titlepage}
\begin{flushright}
SI-HEP-2023-28 \\ 
P3H-23-092\\
Nikhef-2023-020 \\[0.2cm]
\today
\end{flushright}

\vspace{1.2cm}
\begin{center}
\boldmath
{\Large\bf Inclusive Semileptonic $b \to c \ell \bar{\nu}$ Decays  \\[1mm] 
to Order $1/m_b^5$}
\unboldmath
\end{center}

\vspace{0.5cm}
\begin{center}
{\sc Thomas Mannel, Ilija S. Milutin}   \\[2mm]
   Theoretische Physik 1, Center for Particle Physics Siegen \\
   Universit\"at Siegen,  D-57068 Siegen, Germany \\[5mm] 
{\sc K. Keri Vos} \\[2mm] 
{Gravitational Waves and Fundamental Physics (GWFP), \\
Maastricht University, Duboisdomein 30, NL-6229 GT Maastricht, the Netherlands} \\ and \\
{Nikhef, Science Park 105, NL-1098 XG Amsterdam, the Netherlands} 
\end{center}

\vspace{0.8cm}
\begin{abstract}
\vspace{0.2cm}\noindent
Inclusive semileptonic $B\to X_c \ell\bar{\nu}$ decays can be described in the Heavy Quark Expansion (HQE) and allow for a precision determination of the CKM element $|V_{cb}|$. We calculate the terms of $1/m_b^5$ and derive a ``trace formula'' which allows for the computation of the decay rate and kinematic moments of the spectrum up to this order in the HQE. We focus specifically on the reparametrization invariant (RPI) dilepton invariant mass $q^2$ moments of the spectrum, which depend on a reduced set of HQE parameters. We find 10 RPI HQE parameters at $1/m_b^5$. At this order, ``intrinsic charm'' (IC) contributions proportional to $1/(m_b^3m_c^2)$ enter, which are numerically expected to be sizeable. Using the ``lowest-lying state saturation ansatz'' (LLSA), we estimate the size of these contributions. Within this approximation, we observe a partial cancellation between the IC and the ``genuine'' $1/m_b^5$ contributions, resulting in a small overall contribution.  
\end{abstract}

\end{titlepage}

\newpage
\pagenumbering{arabic}
\section{Introduction}
Heavy-quark methods have established themselves as indispensable tools in heavy flavor physics. For inclusive decays, the Heavy Quark Expansion (HQE) has been developed 
to the extend that one can obtain precision predictions. The HQE relies on a systematic expansion in powers of $\Lambda_{\rm QCD} / m_Q$, where $m_Q$ is the mass of the heavy quark and $\Lambda_{\rm QCD}$ is the scale induced by the running QCD coupling constant. One of the prime examples is the determination of the CKM parameter $V_{cb}$
from inclusive $b \to c \ell \bar{\nu}$ transitions, which has reached a relative precision of about one to two percent \cite{Bordone:2021oof,Bernlochner:2022ucr, Finauri:2023kte}. 

The HQE for inclusive semileptonic $b \to c$ transitions is set up as an operator product expansion (OPE) using the 
full QCD heavy quark states. The HQE parameters, forward matrix elements of local operators, are the non-perturbative inputs of the order $\Lambda_{\rm QCD}$ raised to the appropriate power according to the dimension of the HQE parameter. The coefficients of the HQE parameters can be calculated in 
perturbation theory, such that the HQE is a combined expansion in $\alpha_s$ and $\Lambda_{\rm QCD} / m_Q$. 

 For the inclusive $B\to X_c \ell\bar\nu$ decay, the leading term is known to $\alpha_s^3$ for the total rate \cite{Fael:2020tow} and up to $\alpha_s^2$ for the kinematic moments\footnote{The $\alpha_s^2$ corrections to the $q^2$ moments are currently known without a kinematic cut \cite{Fael:2022frj}. Recently, also the $\beta_0 \alpha_s^2$ were calculated \cite{Finauri:2023kte}.}. At $\Lambda_{\rm QCD}^2 / m_Q^2$, two HQE parameters $\mu_\pi^2$
and $\mu_G^2$ enter, which are known up to order $\alpha_s$ \cite{Becher:2007tk,Alberti:2012dn,Alberti:2013kxa}. At $\Lambda_{\rm QCD}^3 / m_Q^3$, the HQE parameters $\rho_D^3$ and $\rho_{LS}^3$ have known coefficients calculated to order $\alpha_s$ \cite{Mannel:2021zzr}. 

Starting at order $\Lambda_{\rm QCD}^4 / m_Q^4$ the number of HQE parameters starts to proliferate, and the coefficients 
are known only at tree level. In general, there are nine independent HQE parameters at order $\Lambda_{\rm QCD}^4 / m_Q^4$, and 
at order $\Lambda_{\rm QCD}^5 / m_Q^5$ there are 18 independent HQE parameters \cite{Mannel:2010wj}. This proliferation is reduced in reparametrization invariant (RPI) observables, like the total rate, which depend on a reduced set of HQE parameters \cite{Mannel:2018mqv, Fael:2018vsp}.    

In this respect, the recently measured dilepton invariant mass $q^2$ moments of the inclusive $B\to X_c \ell \bar{\nu}$ spectrum by the Belle \cite{Belle:2021idw} and Belle II \cite{Belle-II:2022evt} collaborations
have provided new insights. These moments are RPI, unlike other observables like lepton energy and hadronic invariant mass moments. 
The reduction of HQE parameters through RPI could open the way for a full extraction of these HQE elements up to 
$1/m_b^4$ purely from data. A first analysis of $q^2$ moments was done \cite{Bernlochner:2022ucr}, leading to small values for the $1/m_b^4$ elements consistent with zero within uncertainties.

For $B\to X_c$ decays, this expansion is usually set up by fixing the ratio $\rho\equiv m_c^2 / m_b^2$  
of the charm quark mass and the bottom quark mass, where the sensitivity to the infrared pole of the charm mass enters at 
$1/m_b^3$ as $\log \rho$ and at $1/m_b^5 \times 1/\rho = 1/(m_b^3 m_c^2)$. These ``intrinsic charm (IC)'' effects were discussed 
in \cite{Bigi:2009ym, Breidenbach:2008ua} and also more recently for inclusive $D$ meson decays \cite{Fael:2019umf}. Numerically, we have approximately $m_c^2 = m_b \Lambda_{\rm QCD}$, 
which suggests to count $\Lambda_{\rm QCD}^5 / (m_b^3 m_c^2) \sim \Lambda_{\rm QCD}^4 / m_b^4$. Based on this power-counting argument, a full analysis of $1/m_b^4$ would require the inclusion of these terms.

In this paper, we derive these intrinsic charm contributions and simultaneously derive all the $1/m_b^5$ (i.e. the dimension-8) contributions to the HQE for inclusive 
semileptonic $b \to c$ transitions. We derive\footnote{After the publication of this paper, \cite{Finauri:2025ost} pointed out a mistake in our trace formulae at $\mathcal{O}(1/m_b^5)$ as our results did not obey a relation between the axial and tensor terms dictated by gamma matrix relations.  We identified the mistake in our formulae and update our results accordingly in this update. We thank Gael Finauri for pointing this out and for elaborate discussions on this point.} a ``trace formula'' which allows to compute any decay distribution (or moment) 
up to $1/m_b^5$. In addition, we derive the reduced set of RPI operators up to $1/m_b^5$ expanding on \cite{Mannel:2018mqv}. We explicitly study the effect of these higher-order terms on the $q^2$ moments of the spectrum.

Our paper is outlined as follows. In Sec.~\ref{sec:hqe}, we start by setting up
the HQE and identify the HQE parameters as forward matrix elements of local operators up to $1/m_b^4$. In Sec.~\ref{sec:hqemb5}, we determine all the RPI operators up to $1/m_b^5$, and find that there are 10 of these. The trace formula is derived in Sec.~\ref{sec:trace}, which allows us to determine the IC contributions. In Sec.~\ref{sec:pheno} we make a quantitative estimate for the effects of these higher-order terms on $q^2$ moments. To do so, we make use of the ``lowest-lying state saturation ansatz'' (LLSA) as discussed in 
\cite{Heinonen:2014dxa}. Based on this we give an estimate for the impact of the dimension-8 contributions, for the 
intrinsic charm contributions as well as for the full dimension-8 terms. We conclude in Sec.~\ref{sec:conc}. Finally, we collect useful information in the Appendices; conversions to switch between different bases of HQE parameters and details on the derivation of the RPI elements. Furthermore, we provide two ancillary Mathematica notebooks with the expressions for the trace formula and the $q^2$ moments.

\section{HQE and reparametrization invariance}\label{sec:hqe}
We consider the inclusive $B\to X_c\ell\bar{\nu}$ decay:
\begin{align}
    B(p_B)\to X_c(p_X)\,\ell(p_\ell)\,\bar{\nu}(p_\nu)\ ,
\end{align}
where $q\equiv p_\ell+p_\nu$. For semileptonic $b\to c$ decays, the HQE is set up by applying the optical theorem to the correlation function 
of two $b \to c$ weak currents 
\begin{align}\label{eq:Rmunu}
    R_{\mu\nu}(q)&=\int \text{d}^4x\ e^{iq\cdot x} \langle B(v)| T[\bar{b}(x)\Gamma_\mu c(x)\bar{c}(0)\bar{\Gamma}_\nu b(0)] | B(v) \rangle 
    \ ,
\end{align}
where $| B(v) \rangle $ is the $B$ meson state of full QCD moving with the velocity $v = p_B / m_B$ 
and $\Gamma_\mu = \gamma_\mu (1-\gamma_5)$. 
The $b$ quark field is then redefined according to 
\begin{align} \label{b-redef}
    b(x)&=e^{-im_b(v\cdot x)}b_v(x)\ ,
\end{align}
which is equivalent to a decomposition of the $b$-quark momentum into $p_b = m_b v + k$, where $k$ is a residual momentum 
with $k \sim \Lambda_{\rm QCD}$. Expanding then in powers of $k/m_b$ generates the OPE of the time-ordered product in  \eqref{eq:Rmunu}, yielding the HQE for $R_{\mu \nu}$. Technically, this means that the dynamical degrees of freedom 
of the bottom quark are integrated out at some scale $\mu \sim m_b$, leaving us with a static $b$ quarks. In our approach, we integrate out the
bottom and charm quarks at the same scale $\mu \sim m_c \approx m_b$, e.g. at $\mu = \sqrt{m_b m_c}$, while keeping  $\rho= m_c^2/m_b^2$ as a number of order unity. Only the light quarks 
(treated as massless) remain dynamical. 

Symbolically, this leads to 
\begin{align} \label{Rs} 
    {\cal R}_{\mu \nu} (S) &= 
    \int \text{d}^4x\ e^{-i m_b (S \cdot x)}   T[\bar{b}_v(x)\Gamma_\mu c(x)\bar{c}(0)\bar{\Gamma}_\nu b_v(0)] \\ 
    &=\sum\limits_{n=0}^\infty C_{\mu \nu\,\mu_1 ... \mu_n}^{(n)}(S)\otimes \bar{b}_v(iD^{\mu_1})...(iD^{\mu_n})b_v\ ,
    \nonumber
\end{align}
where $S\equiv v-q/m_b$, $\otimes$ denotes the contraction of the Dirac indices, 
and the coefficients $C^{(n)}$ carry mass dimensions $1/m_b^{n+3}$. Taking the forward matrix element 
of \eqref{Rs} yields the desired $1/m_b$ expansion for the total rate as well as for kinematic moments. The matrix 
elements of the operators appearing in \eqref{Rs}, 
$$
\langle B(v) | \bar{b}_v(iD^{\mu_1})...(iD^{\mu_n})b_v | B(v) \rangle 
$$
are decomposed into scalar quantities, which can be expresses in terms of forward matrix elements of scalar operators, 
and which define the HQE parameters.

It has been discussed at length that starting at order $1/m_Q^4$, the number of independent parameters in the HQE 
proliferates, making an extraction of all these parameters from data impossible. However, 
as worked out in \cite{Mannel:2018mqv}, and discussed before in \cite{Luke:1992cs,Manohar:2010sf, Dugan:1991ak, Chen:1993np}, both the OPE as well as the HQE obey reparametrization invariance (RPI). Since the vector $v$ has been introduced artificially 
in \eqref{b-redef}, the expression $\langle B(v) | \mathcal{R}(S) | B(v) \rangle$ cannot depend on $v$, so a reparametrization
transformation $\delta_{\rm RP}: v_\mu  \mapsto v_\mu + \delta v_\mu$ and simultaneously $\delta_{\rm RP}iD_\mu= -m_Q \delta v_\mu$ 
with $v \cdot \delta v = 0$ should leave $\langle B(v) | \mathcal{R}(S) | B(v) \rangle$ invariant. Consequently, $\delta_{\rm RP}$ links different orders in $1/m_Q$ through \cite{Mannel:2018mqv}
\begin{align}\label{eq:rpirel}
    \delta_{\rm{RP}}C_{\mu_1...\mu_n}^{(n)}(S)&=m_Q\ \delta v^\alpha\left(C^{(n+1)}_{\alpha\mu_1...\mu_n}(S)+C^{(n+1)}_{\mu_1\alpha...\mu_n}(S)+...+C^{(n+1)}_{\mu_1...\mu_n\alpha}(S)\right)\ ,
\end{align}
which leads to a reduction of the number of independent parameters for RPI quantities.

Although we have derived all relations to order $1/m_b^5$ for the general case, we will restrict our discussion
to the case of RPI observables. For the case at hand, this means the total rate and the moments of the leptonic invariant mass $q^2$. Up to $1/m_b^4$, we define \cite{Mannel:2018mqv,Fael:2018vsp},
 \begin{align}
    2m_B\mu_3&=\langle \bar{b}_v b_v\rangle\ ,\nonumber \\
    2m_B\mu_G^2&=\langle \bar{b}_v (iD_\alpha)(iD_\beta)(-i\sigma^{\alpha\beta})b_v\rangle\ ,\nonumber\\
    2m_B\tilde{\rho}_D^3&=\frac{1}{2}\langle \bar{b}_v\Big[(iD_\mu),\Big[\Big((ivD)+\frac{1}{2m_b}(iD)^2\Big),(iD^\mu)\Big]\Big]b_v\rangle\ ,\nonumber\\
    2m_Br_G^4&=\langle \bar{b}_v[(iD_\mu),(iD_\nu)][(iD^\mu),(iD^\nu)]b_v\rangle\ ,\nonumber\\
    2m_Br_E^4&=\langle \bar{b}_v[(ivD),(iD_\mu)][(ivD),(iD^\mu)]b_v\rangle\ ,\nonumber\\
    2m_Bs_B^4&=\langle \bar{b}_v[(iD_\mu),(iD_\alpha)][(iD^\mu),(iD_\beta)](-i\sigma^{\alpha\beta})b_v\rangle\ ,\nonumber\\
    2m_Bs_E^4&=\langle \bar{b}_v[(ivD),(iD_\alpha)][(ivD),(iD_\beta)](-i\sigma^{\alpha\beta})b_v\rangle\ ,\nonumber\\
    2m_Bs_{qB}^4&=\langle\bar{b}_v[(iD_\mu),[(iD^\mu),[(iD_\alpha),(iD_\beta)]]](-i\sigma^{\alpha\beta})b_v\rangle\ ,
    \label{eq:mb4el}
\end{align}
where we have introduced the notation $\langle \bar{b}_v\,...\,b_v\rangle\equiv\langle B(v)|\bar{b}_v\,...\,\bar{b}_v|B(v)\rangle$, and where $\gamma^\mu\gamma^\nu=g^{\mu\nu}+(-i\sigma^{\mu\nu})$. We also note here that compared to its standard definition, $\rho_D^3$ is redefined to include higher-order terms in the $1/m_b$ expansion. The non-RPI matrix elements, required to describe for example the lepton energy spectrum are listed in Appendix~\ref{sec:ap_conversion}, as well as a conversion to the $iD^\perp$ basis used in \cite{Mannel:2010wj, Gambino:2016jkc}. In \cite{{Mannel:2018mqv}} these parameters were written in terms of chromoelectric ($\vec{E}$) and chromomagnetic 
($\vec{B}$) fields, giving some physical intuition on the meaning of these expressions. 

We end the review of the dimension-seven operators by making a remark 
concerning the the operators involving symmetrized products of color octets, such as 
$r_G^4$ and $r_E^4$. At tree level, this involves 
\begin{equation}
\{ T^a, T^b \} = \frac{1}{3}\delta^{ab} + d^{abc} T^c \, . 
\end{equation}
However, beyond tree level the color singlet and the color octet contributions become independent 
operators \cite{Kobach:2017xkw} and thus will lead to additional HQE parameters. Nevertheless, defining the 
matrix elements as in \eqref{eq:mb4el} will be correct up to corrections of order $\alpha_s (m_b)$.

\section{HQE parameters up to \boldmath $1/m_b^5$ \unboldmath}\label{sec:hqemb5}
The HQE parameters at $1/m_b^5$ have been listed in \cite{Mannel:2010wj} for the general case at tree level. For completeness we also give these operators in Appendix \ref{sec:ap_conversion}. 

In order to determine the number of RPI parameters, we follow the construction outlined in \cite{Mannel:2018mqv}. Starting from \eqref{eq:rpirel}, this requires writing down all possible tensor decomposition of the $C$ coefficients. We discuss this derivation in detail in Appendix~\ref{sec:ap_rpiderivation}. We obtain in total 10 RPI parameters:
\begin{align}
    2m_BX_1^5&=\langle \bar{b}_v\Big[(ivD),[(ivD),(iD_\mu)]\Big][(ivD),(iD^\mu)]b_v\rangle\ \nonumber,\\
      2m_BX_2^5&=\langle \bar{b}_v\Big[ (ivD),[(iD_\mu),(iD_\nu)]\Big][(iD^\mu),(iD^\nu)]b_v\rangle\ , \nonumber\\
      2m_BX_3^5&= \langle \bar{b}_v \Big[(iD_\mu),[(ivD),(iD_\nu)][(iD^\mu),(iD^\nu)]\Big]b_v\rangle\ 
 \nonumber\\
     2m_BX_4^5 &=\langle \bar{b}_v \Big[(iD_\mu),\Big[(iD_\nu),\Big[(iD^\mu),[(ivD),(iD^\nu)]\Big]\Big]\Big]b_v\rangle\ , \label{eq:Xidef4}
\end{align}
for the spin-singlet contributions. In addition, we have 6 more operators which are spin-triplets:  
\begin{align}
    2m_BX_5^5&=\langle \bar{b}_v\Big[(ivD),[(ivD),(iD_\alpha)]\Big][(ivD),(iD_\beta)](-i\sigma^{\alpha\beta})b_v\rangle\ ,\nonumber\\
       2m_BX_6^5&=\langle \bar{b}_v\Big[(ivD),[(iD_\mu),(iD_\alpha)]\Big][(iD^\mu),(iD_\beta)](-i\sigma^{\alpha\beta})b_v\rangle\ ,\nonumber \\
       2m_BX_7^5&=\langle \bar{b}_v\Big[(iD_\mu),[(ivD),(iD_\alpha)]\Big][(iD^\mu),(iD_\beta)](-i\sigma^{\alpha\beta})b_v\rangle\ , \nonumber\\
    2m_BX_8^5&=\langle \bar{b}_v\Big[(iD_\mu),[(ivD),(iD_\alpha)][(iD^\mu),(iD_\beta)]\Big](-i\sigma^{\alpha\beta})b_v\rangle\ ,  \nonumber\\
    2m_BX_9^5&=\langle \bar{b}_v\Big[(iD_\mu),[(ivD),(iD^\mu)]\Big][(iD_\alpha),(iD_\beta)](-i\sigma^{\alpha\beta})b_v\rangle\ , \nonumber\\
    2m_BX_{10}^5&=\langle\bar{b}_v\Big[(iD_\mu),[(ivD),(iD^\mu)][(iD_\alpha),(iD_\beta)]\Big](-i\sigma^{\alpha\beta})b_v\rangle\ ,  \label{eq:Xidef10}
\end{align}
These RPI parameters are specific linear combinations of the general list given in \cite{Mannel:2010wj}. In Appendix~\ref{sec:ap_conversion}, we list the relation between the RPI $X_i^5$ operators and the full basis. We note that these operators can be expressed in terms of gluon fields and their derivatives as done for the $1/m_b^4$ operators in \cite{Mannel:2018mqv}. Here, we do not give these relations since the differences between QED and QCD become more involved at higher orders. We also point out again that
we deal here with tree level only, meaning that the remark made in Sec.~\ref{sec:hqe} applies also 
at dimension eight. 

Finally, we stress that the equation of motion 
\begin{equation}\label{eq:eom}
\left((ivD) + \frac{1}{2m_b} (iD)^2\right) b_v = - \frac{1}{2m_b} (\sigma\cdot G)\, b_v \ ,
\end{equation}
with 
\begin{equation}
    \sigma \cdot G \equiv (-i\sigma_{\mu\nu}) iD^\mu iD^\nu \ .
\end{equation}
as well as $\sigma \cdot G $ itself is RPI. This already lead to the ``RPI completed'' expression for 
$\rho_D^3 \to \tilde{\rho}_D^3$ shown in \eqref{eq:mb4el}, and likewise we will have the ``RPI completed''
expressions for $r_E^4$ and $s_E^4$, which now also contain $1/m_b^5$ (and higher) terms:
\begin{align}
 2m_B\tilde{r}_E^4&=\langle \bar{b}_v\Big[\Big((ivD)+\frac{1}{2m_b}(iD)^2\Big),(iD_\mu)\Big]\Big[\Big((ivD)+\frac{1}{2m_b}(iD)^2\Big),(iD^\mu)\Big]b_v\rangle\ ,\label{eq:RPIcompletion_rEsE} \\ \nonumber 
    2m_B\tilde{s}_E^4&=\langle \bar{b}_v\Big[\Big((ivD)+\frac{1}{2m_b}(iD)^2\Big),(iD_\alpha)\Big]\Big[\Big((ivD)+\frac{1}{2m_b}(iD)^2\Big),(iD_\beta)\Big](-i\sigma^{\alpha\beta})b_v\rangle \ . 
\end{align}
Combined, this gives the full list of RPI operators up to $1/m_b^5$.

\section{\boldmath $B\to X_c \ell \bar{\nu}$ at $1/m_b^5$ \unboldmath}\label{sec:trace}
In this section, we outline the calculation to obtain the total rate and kinematic moments for the 
inclusive decay $B \to X_c \ell \bar{\nu}$. We start from $\mathcal{R}(S)$ in \eqref{Rs}, which is 
related to the hadronic tensor of  the $B\to X_c \ell\bar{\nu}$ transition via
\begin{equation}
    W(v,q) = -\frac{1}{\pi}{\rm Im}  \langle B(v) | {\cal R}(S) | B(v)\rangle  \ .\label{optical}
\end{equation} 

Our goal is to formulate a ``trace formula'' \cite{Mannel:2010wj,Mannel:1994kv,Dassinger:2006md} 
to compute the observables for $B\to X_c \ell \bar{\nu}$ 
including all terms up to dimension-8 operators at tree level. This is achieved by observing that 
the time-ordered product in \eqref{eq:Rmunu} can be written in terms of the ``external field propagator'' 
of the charm quark as 
\begin{eqnarray} \label{eq:HQE1} 
&& \int \text{d}^4x\ e^{-im_b(S\cdot x)}  T[\bar{b}_v(x)\Gamma_\mu c(x)\bar{c}(0)\bar{\Gamma}_\nu b_v(0)] = 
\bar{b}_v(0) \Gamma_\mu  \left(\frac{1}{\slashed{Q} + i \slashed{D} - m_c}  \right) \bar{\Gamma}_\nu b_v (0) \nonumber \\ 
&& = \bar{b}_v(0) \Gamma_\mu \left(\frac{1}{\slashed{Q} - m_c} \right) \sum_{k = 0}^\infty 
\left[ (i\slashed{D}) \left(\frac{1}{\slashed{Q} - m_c} \right)   \right]^k  \bar{\Gamma}_\nu b_v(0) \nonumber \\ 
&& = \left\{ \Gamma_\mu \left(\frac{1}{\slashed{Q} - m_c} \right) \bar{\Gamma}_\nu \right\}_{\alpha \beta}
\bar{b}_{v,\alpha}  b_{v,\beta}  \nonumber  \\ 
&& \quad + \left\{ \Gamma_\mu \left(\frac{1}{\slashed{Q} - m_c} \right)   
 \gamma^{\rho_1} \left(\frac{1}{\slashed{Q} - m_c} \right)   \bar{\Gamma}_\nu \right\}_{\alpha \beta}
\bar{b}_{v,\alpha} (iD_{\rho_1})   b_{v,\beta}   \nonumber \\ 
&& \quad + \left\{ \Gamma_\mu \left(\frac{1}{\slashed{Q} - m_c} \right) \gamma^{\rho_1}
\left(\frac{1}{\slashed{Q} - m_c} \right) \gamma^{\rho_2} \left(\frac{1}{\slashed{Q} - m_c} \right) 
\bar{\Gamma}_\nu \right\}_{\alpha \beta} \bar{b}_{v,\alpha} (iD_{\rho_1}) (iD_{\rho_2})  b_{v,\beta}
\nonumber \\ 
&& \quad +\, \ldots
\end{eqnarray}
where $Q\equiv m_bv-q$. Taking the forward  matrix element of this expression yields 
\begin{eqnarray}\label{trace_expansion} 
&& \langle B(v) | \bar{b}_v(0) \Gamma_\mu \left(\frac{1}{\slashed{Q} - m_c} \right) \sum_{k = 0}^\infty 
\left[ (i\slashed{D}) \left(\frac{1}{\slashed{Q} - m_c} \right)   \right]^k  \bar{\Gamma}_\nu b_v(0) | B(v) \rangle  \nonumber \\ 
&& = {\rm Tr} \left[ \left\{ \Gamma_\mu \left(\frac{1}{\slashed{Q} - m_c} \right) \bar{\Gamma}_\nu \right\} {\cal M}^{(3)} \right]
  \nonumber  \\ 
&& \quad + {\rm Tr} \left[  \left\{ \Gamma_\mu \left(\frac{1}{\slashed{Q} - m_c} \right)   
 \gamma^{\rho_1} \left(\frac{1}{\slashed{Q} - m_c} \right)   \bar{\Gamma}_\nu \right\} 
{\cal M}^{(4)}_{\rho_1} \right]   \nonumber \\ 
&& \quad + {\rm Tr} \left[  \left\{ \Gamma_\mu \left(\frac{1}{\slashed{Q} - m_c} \right) \gamma^{\rho_1}
\left(\frac{1}{\slashed{Q} - m_c} \right) \gamma^{\rho_2} \left(\frac{1}{\slashed{Q} - m_c} \right) 
\bar{\Gamma}_\nu \right\} {\cal M}_{\rho_1 \rho_2}^{(5)} \right]
\nonumber \\ 
&& \quad +\, \ldots 
\end{eqnarray}
where the hadronic matrix elements are given by the Dirac matrices
\begin{eqnarray}
\{ {\cal M}^{(3)} \}_{\beta \alpha} &=& \langle  \bar{b}_{v,\alpha}  b_{v,\beta}   \rangle\ , \nonumber\\
\{ {\cal M}^{(4)}_{\rho_1} \}_{\beta \alpha} &=& 
\langle   \bar{b}_{v,\alpha} (iD_{\rho_1}) b_{v,\beta}   \rangle\ ,\nonumber \\
\{ {\cal M}^{(5)}_{\rho_1 \rho_2} \}_{\beta \alpha} &=& 
\langle  \bar{b}_{v,\alpha} (iD_{\rho_1}) (iD_{\rho_2})  b_{v,\beta}  \rangle\ ,\nonumber\\
 ... \label{traceformulaslist}
\end{eqnarray} 
In order to compute to the desired order, we start at the highest order corresponding to $n=k+3$. Since we 
neglect all higher-order terms, we can compute this matrix element in the static limit, which means 
    \begin{equation}
        \langle \bar{b}_{v, \alpha} (iD_{\mu_1})...(iD_{\mu_k})b_{v,\beta}\rangle 
        = \{ {\cal M}^{(k+3)}_{\mu_1... \mu_k} \}_{\beta \alpha} \hspace{0.3cm} \mbox{with} \hspace{0.3cm}  
        {\cal M}^{(k+3)}_{\mu_1...\mu_k} = {\bf 1} A_{\mu_1...\mu_k}+ s_\lambda B^\lambda_{\mu_1...\mu_k} \ ,
    \end{equation}
where $P_+ = (1+\slashed{v})/2$ is the projector on the ``large'' components of a Dirac spinor and  $s_\lambda= P_+ \gamma_\lambda \gamma_5 P_+$ corresponds to the three Pauli matrices in the rest frame $v = (1, \vec{0})$. 

Matrix elements of lower dimension are then obtained with an iterative process taking into account 
all possible Dirac structures: 
    \begin{equation}
        \langle \bar{b}_{v,\alpha} (iD_{\mu_1})...(iD_{\mu_l}) b_{v,\beta}\rangle = 
        \{ {\cal M}^{(l+3)}_{\mu_1 ... \mu_l} \}_{\beta \alpha} \quad \mbox{with} \quad 
        {\cal M}^{(l+3)}_{\mu_1 ... \mu_l} = 
        \sum_i \Gamma^i A^i_{\mu_1...\mu_{l}}\ .
    \end{equation}
    where the sum now runs over the complete set of Dirac matrices 
    $\Gamma_i = \{{\bf 1}, \gamma_5, \gamma_\mu, \gamma_5 \gamma_\mu, \sigma_{\mu \nu} \} $. 
The tensors $A^i$ are finally expressed in terms of the HQE parameters.

The resulting trace formulae including terms up to $1/m_b^5$ in the full basis are given in a Mathematica notebook added as an ancillary file (see Appendix \ref{sec:ap_traceform} for more details). We note that these formulae were already derived in \cite{Mannel:2010wj}, but there were not publicly available\footnote{We thank the authors for providing us with a Mathematica notebook containing their formulae. We reproduce their results up to $1/m_b^4$. At $1/m_b^5$, we found a few mistakes in their derivation which we correct in our trace formula.}. Furthermore, in this updated version, we correct for a mistake found in our old trace formula, which was identified in \cite{Finauri:2025ost}, resulting in a difference proportional to $-X_8^5+\frac{1}{2}X_{10}^5$. This only results in changes in the $X_8^5$ and $X_{10}^5$ terms in Appendices \ref{sec:ap_traceform}, \ref{sec:ap_totalrate} and \ref{sec:ap_q2moments}, w.r.t.~the previous version of this work. The conclusions and other results of this work are not altered.

Inserting the expression found with the trace formulae from  \eqref{trace_expansion} and taking the imaginary part of the hadronic correlator according to \eqref{optical}, allows us to find the functions $W_i(q^2, v\cdot q)$ of the Lorentz decomposition of $W$:
\begin{align}
    W_{\mu\nu}&=-g_{\mu\nu}W_1+v_\mu v_\nu W_2-i\varepsilon_{\mu\nu\rho\sigma}v^\rho q^\sigma W_3+q_\mu q_\nu W_4+(q_\mu v_\nu+q_\nu v_\mu)W_5\ .
\end{align}
From this, the triple differential rate can be obtained and finally also moments of the kinematic distributions. 

These moments are normalized integrated quantities defined by 
\begin{equation}
    \langle (O)^n \rangle_{\mathrm{cut}} =
    \int_{\mathrm{cut}}
    (O)^n 
    \frac{\mathrm{d} \Gamma}{\mathrm{d} O} \, \mathrm{d} O
    \Bigg/
    \int_{\mathrm{cut}}
    \frac{\mathrm{d} \Gamma}{\mathrm{d} O} \, \mathrm{d} O\ ,
\end{equation}
where $O$ are observables like $O= E_\ell, M_X^2, q^2,..$. The subscript ``cut'' generically denotes some restriction in the 
lower integration limit. In the following, we discuss only the $q^2$ moments in detail, as these can be expressed in terms of the 10 $X_i^5$ RPI operators. We consider centralized moments defined through
\begin{align*}
    q_1(q^2_\mathrm{cut}) &= \langle q^2 \rangle_{q^2 \ge q^2_\mathrm{cut}}\ , & 
    q_n (q^2_\mathrm{cut}) &= \Big\langle (q^2 - \langle q^2 \rangle )^n \Big\rangle_{q^2 \ge q^2_\mathrm{cut}}
    \text{ for } n\ge2\ .
\end{align*}
We furthermore also define
 \begin{align}\label{eq:rstar}
     R^*(\hat{q}^2_{\rm{cut}})\equiv 
    \int_{\hat{q}^2_\mathrm{cut}}^{(1-\sqrt{\rho})^2}
    \frac{\mathrm{d} \Gamma}{\mathrm{d} \hat{q}^2} \, \mathrm{d} \hat{q}^2
    \Bigg/
   \int_{0}^{(1-\sqrt{\rho})^2}
    \frac{\mathrm{d} \Gamma}{\mathrm{d} \hat{q}^2} \, \mathrm{d} \hat{q}^2\ .
 \end{align}
For completeness, we give the total rate in terms of the RPI parameters in
Appendix~\ref{sec:ap_totalrate}.  We note that this expression differs from the one 
given in \cite{Gambino:2016jkc}, where the $1/m_b^{4,5}$ parameters were extracted 
from moments of the lepton energy and $M_X$ spectrum. Using the conversion between 
the different bases in Appendix~\ref{sec:ap_conversion}, we find that the rate presented 
in \cite{Gambino:2016jkc} is not RPI. We also give in Appendix~\ref{sec:ap_q2moments}  
for the first time the $q^2$ moments up to $1/m_b^5$. The expressions for the first 
four moments including a $q^2$-cut are also given as an ancillary file (see Appendix \ref{sec:ap_traceform} for more details).

\subsection{Intrinsic charm contributions}
As discussed in the introduction, the dimension-8 contributions 
contain terms involving negative powers of $m_c^2$. In fact starting at $1/m_b^3$ the HQE, where 
the bottom and the charm quark are integrated out simultaneously, exhibits an infrared sensitivity to the 
charm-quark mass, and the effects related to this are usually called ``intrinsic charm'' (IC) \cite{Bigi:2009ym,Breidenbach:2008ua}.  
This IR sensitivity to the charm mass comes in through the phase space integration over $v\cdot Q$. The integrands are singular staring at 
dimension six, i.e. $1/m_b^3$, which finally results in $\log \rho$ 
terms in the total rate and the moments 
of $B\to X_c$.  However, at higher dimensions $n$, we even 
pick up power-like singularities for $m_c\to 0$ resulting in a contribution to the total rate of
\begin{equation}
\Gamma_n \simeq \frac{1}{m_b^3} \left(\frac{1}{m_c^2}\right)^{(n-6)/2} \ , 
\end{equation}
where $n=8,10,12,...$. Including $\alpha_s$ corrections will also introduce odd powers 
of $1/m_c$, but here we consider only tree-level contributions.

\begin{figure}[t]
\centerline{\includegraphics[scale=0.33]{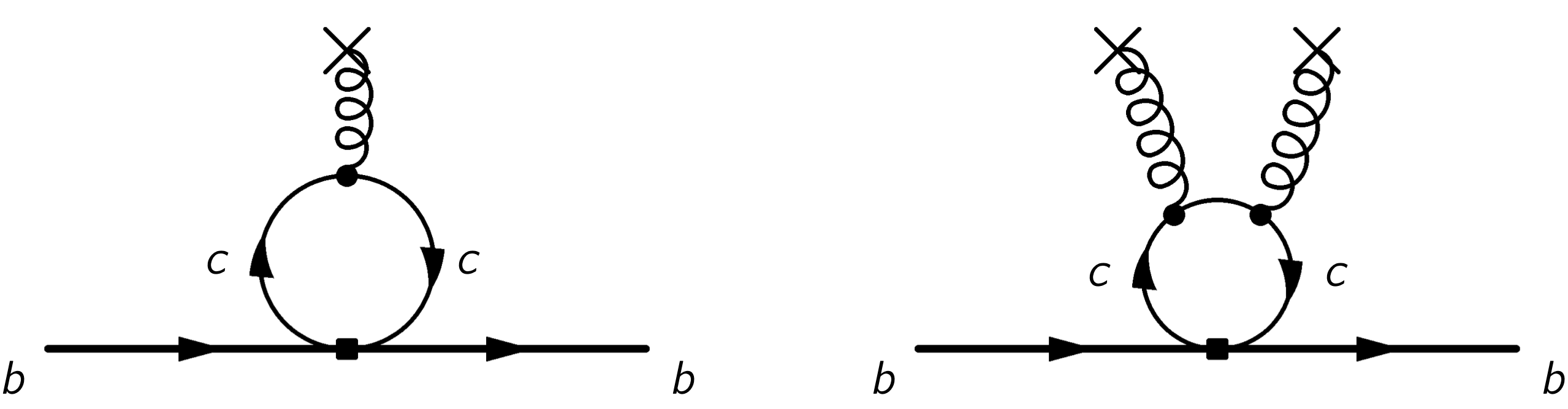}}
\caption{Feynman Diagrams for the ``intrinsic charm'' contributions.}
\label{fig:IC}
\end{figure}

Following \cite{Bigi:2009ym}, 
we can extract the relevant expressions by considering the charm propagator in an external field according to 
the Feynman Diagrams shown in Fig.~\ref{fig:IC}. Expanding this expressions to $1/m_c^2$, one finds
\begin{eqnarray}
\langle \bar{c}_\alpha  \gamma_\nu \gamma_5  c_\beta \rangle_A &=& 
\frac{1}{48 \pi^2 m_c^2} \left(
    2 \left\{ \left[ D_\kappa \, , \, G^{\kappa \lambda}  \right] \, , \, \tilde{G}_{\mu \lambda} \right\} 
    +  \left\{ \left[ D_\kappa \, , \,  \tilde{G}_{\mu \lambda} \right] \, , \,  G^{\kappa \lambda}  \right\} 
    \right)_{ \beta \alpha} + \, ...\label{IC_gamma5}
 \\
\langle \bar{c}_\alpha  \gamma_\nu    c_\beta \rangle_A &=& 
\frac{2}{3} \frac{1}{16 \pi^2} \ln\left(\frac{m_b^2}{m_c^2} \right) 
[D^\kappa \, , \, G_{\kappa \nu}]_{\beta \alpha} \nonumber \\ 
&& +
\frac{i}{240 \pi^2 m_c^2} \Big( 13 \left[ D^\kappa \, , \, \left[ G_{\lambda \nu } \, , \, G^{\lambda \kappa} 
\right] \right] + 8 i  \left[ D^\kappa \, , \,  \left[ D^\lambda \, , \, \left[ D_\lambda \, , \, G_{\kappa \nu} \right] 
\right] \right] \label{IC_gamma} \\ 
&& \qquad \qquad \quad  
 - 4 i  \left[ D^\lambda \, , \,  \left[ D^\kappa \, , \, \left[ D_\lambda \, , \, G_{\kappa \nu} \right] 
\right] \right]
\Big)_{ \beta \alpha} + \, ... \nonumber 
\end{eqnarray} 
where the subscript $A$ denotes that the propagator is to be take in an external gluon field $A$, 
$\alpha$ and $\beta$ are color indices, and we have inserted $m_b$ as the proper UV cut-off in the 
first term of (\ref{IC_gamma}), which generates the well-known $\log (m_c^2 / m_b^2)$ term in the 
coefficient of $\rho_D^3$. 

Inserting this into \eqref{eq:HQE1}, we end up with dimension-eight operators proportional to 
$1/(m_b^3  m_c^2)$. Note that the IC contribution is RPI, so it can be expressed in terms of our 
basis of RPI operators $X_1^{5},... ,X_{10}^{5}$ defined in (\ref{eq:Xidef4},\ref{eq:Xidef10}):
\begin{align}
    X_{\rm{IC}}^{5} 
    &\equiv-24X_2^5+78X_3^5-12X_4^5+10X_7^5-20X_8^5+5X_9^5+5X_{10}^5\ ,\label{eq:XIC_def}
\end{align} 
Alternatively, having the full trace formula available, we can also easily identify this 
combination from the total rate, given in Appendix \ref{sec:ap_totalrate}, by identifying 
the HQE parameters proportional to $1/\rho$.  We explicitly checked that the same combination
$X_{\rm{IC}}^{5} $ of dimension-eight HQE parameters $X_{\rm{IC}}^{5}$ also describes the 
intrinsic charm contributions proportional to $1/(m_b^3m_c^2)$ in the centralized moments 
$q_{1-4}$ as required by RPI.

\section{Phenomenological implications}\label{sec:pheno}
In the remainder of the paper, we will discuss the phenomenology of the dimension-8 contributions to the 
HQE. Since the number of HQE parameters - even for the reduced set using RPI - is too large to extract 
them from the data, we shall employ ``lowest-lying state saturation ansatz'' (LLSA) to obtain an idea of the size of the $1/(m_b^3 m_c^2)$ and 
$1/m_b^5$ terms of the HQE. Overall it turns out that both the IC contributions of order $1/(m_b^3 m_c^2)$
and the ``genuine'' $1/m_b^5$ pieces are sizable and of similar magnitude, although the IC contribution 
should be parametrically larger. However, the two contributions enter with different sign, which leads 
overall to a small contribution of the dimension-8 operators. We shall discuss this 
issue in Subsec.~\ref{sec:IC}.

\subsection{Estimation of the matrix elements}
Before we proceed to discuss phenomenological implications of the decay $B \to X_c \ell \bar{\nu}$, we will try to estimate the size 
of the HQE parameters at dimension-8. To do so, we use the ``lowest-lying state saturation ansatz'' (LLSA) 
which has been elaborated upon in \cite{Heinonen:2014dxa}. The starting point is to introduce a fictitious 
heavy quark $Q$ with $m_Q \gg m_b$ for which we can derive the sum rule  
\begin{eqnarray}
&& \sum_n \frac{i (2 \pi)^3 \delta^3 (p_n^\perp)}{\omega - \epsilon_n + i \varepsilon} 
\langle B(p_B) | \bar{b}_v [(iD_{\mu_1}^\perp) ... (iD_{\mu_k}^\perp)] Q_v | n \rangle \, 
\langle n | \bar{Q}_v [(iD_{\nu_1}^\perp) ... (iD_{\nu_l}^\perp)] b_v | B(p_B) \rangle 
\nonumber \\ 
&& \qquad = \langle B(p_B) | \bar{b}_v [(iD_{\mu_1}^\perp) ... (iD_{\mu_k}^\perp)] 
\left(\frac{i}{\omega + ivD + i \varepsilon} \right) 
  [(iD_{\nu_1}^\perp) ... (iD_{\nu_l}^\perp)] b_v | B(p_B) \rangle\ , 
  \label{LLSASR} 
\end{eqnarray} 
where the superscript $\perp$ denotes the ``spatial'' components of a vector, 
$$
a^\perp_\mu \equiv a_\mu - v_\mu (av) = (g_{\mu \nu} - v_\mu v_\nu) a^\nu \equiv g_{\mu \nu}^\perp a^\nu \ ,
$$
$Q_v$ is the static field 
of the heavy quark, while $b_v$ is still the field of full QCD with the definition 
$$
b(x) = e^{-im_b (v\cdot x)} b_v(x) \ ,
$$
to remove the large part of the quark momentum $m_b v$. 

This sum rule can now be expanded in powers of $1/\omega$ to generate the matrix elements defining the 
HQE parameters. The LLSA is to truncate the sum over all intermediate states on the left-hand 
side after the first non-vanishing terms. 

Following \cite{Heinonen:2014dxa},
we anchor the LLSA by using (\ref{LLSASR}) with $k = 1$ and $l=1$. This requires to consider 
the matrix elements 
$$
\langle B(p_B) | \bar{b}_v (iD_\mu^\perp)  Q_v | n \rangle  
$$
for the tower of states $|n \rangle$. The lowest lying states are the two spin-symmetry 
doublets of orbitally excited $\ell = 1$ states, consisting of $(0^+,1^+)_{j_{\rm light} = 1/2}$ and
$(1^+,2^+)_{j_{\rm light} = 3/2}$ states, where $j_{\rm light}$ denotes the total angular momentum of the 
light degrees of freedom. The two states within the doublet are degenerate 
(in the $m_Q \to \infty$ limit), and the two doublets have excitation energies  
$\epsilon_{1/2}$ and $\epsilon_{3/2}$ relative to the ground state. 

For the lowest term in the $1/\omega$ expansion of (\ref{LLSASR}) with $k = 1$ and $l=1$ 
the left hand side becomes $(\mu_\pi^2)^\perp$ or $(\mu_G^2)^\perp$ (see below), depending on which 
matrix $\Gamma$ is inserted, 
while the sum on the left-hand side is truncated 
after the contributions of the two spin-symmetry doublets discussed above. This allows us to fix the 
values of the two matrix elements 
$$
\langle B(p_B) | \bar{b}_v (iD_\mu^\perp)  Q_v | \ell = 1, j_{\rm light} = 1/2 \rangle 
\quad \mbox{and} \quad \langle B(p_B) | \bar{b}_v (iD_\mu^\perp)  Q_v 
| \ell = 1, j_{\rm light} = 3/2 \rangle 
$$
in terms of $(\mu_\pi^2)^\perp$ or $(\mu_G^2)^\perp$. 
Inserting more $\perp$-derivatives, i.e. $k>1$ and $l>1$ and expanding to higher orders 
in $1/\omega$ one then can relate $\rho_D^3$ ($k = 1$,  $l=1$ and expanding to $1/\omega$)
and all higher-order HQE elements to the $1/m_b^2$ HQE elements $(\mu_\pi^\perp)^2$ and $(\mu_G^\perp)^2$ and the excitation energies $\epsilon_{1/2}$ and 
$\epsilon_{3/2}$ of the orbitally excited states with $j_{\rm light} = 1/2$ and $j_{\rm light} = 3/2$, respectively. In Appendix \ref{sec:ap_LLSA}, we list the LLSA approximations for all the RPI HQE elements up to $1/m_b^5$. Similar expressions can be found in \cite{Heinonen:2014dxa} for the full basis of $1/m_b^5$ elements.

To obtain numerical estimates for the HQE parameters, we take the excitation energies from the decay spectrum \cite{Heinonen:2014dxa}
\begin{equation}
    \epsilon_{1/2}= 0.390\ {\rm GeV}\ , \quad\quad \epsilon_{3/2} = 0.476\ {\rm GeV}\ .
\end{equation}
In addition, the HQE parameters are known from a global analysis of the $B\to X_c \ell \bar{\nu}$ spectrum \cite{Bordone:2021oof}\footnote{Taking the extracted HQE parameters from the recent fit from \cite{Finauri:2023kte} does not change our conclusions, but only changes the values for the HQE parameters in Table \ref{tab:inputs} by approximately $0-10\%$. }
\begin{equation}
    (\mu_\pi^2)^\perp = 0.477\ {\rm GeV}^2 \ , \quad\quad (\mu_G^2)^\perp = 0.306\  {\rm GeV}^2\ ,
\end{equation}
which are defined as
\begin{align}
    2m_B(\mu_\pi^2)^\perp&=-\langle \bar{b}_v\, (iD_\mu)\, (iD_\nu)\, b_v\rangle g^{\mu\nu}_\perp\nonumber\ ,\\
    2m_B(\mu_G^2)^\perp&=\langle \bar{b}_v\, (iD_\alpha)\, (iD_\beta)\, (-i\sigma_{\mu\nu})\,b_v\rangle g_\perp^{\mu\alpha}g_\perp^{\nu\beta}\ .
\end{align}

Using these values and the LLSA expressions in Appendix \ref{sec:ap_LLSA}, we find the approximations for the HQE parameters presented in Table \ref{tab:inputs}.  We do not show any 
uncertainty range, since we currently do not have a way to estimate the quality of the LLSA. Instead, we only use these numerical values to get an estimate for the size and sign of the contributions as was done also in \cite{Gambino:2016jkc}. We note that, we can also take the $n^{th}$ root of the absolute values of the HQE parameters, which yields values of order $\Lambda_{\rm{QCD}}$ as expected. An exception to this is $\sqrt[\leftroot{-2}\uproot{2}4]{|s_{qB}^4|}\approx 1$ GeV.

For the intrinsic charm contribution $X_{\rm{IC}}^{5}$ defined in \eqref{eq:XIC_def}, we find 
\begin{align}
    X_{\rm{IC}}^{5}=&\ \frac{10}{9}\left[4\epsilon_{1/2}\Big(\Big((\mu_G^2)^\perp\Big)^2-7(\mu_G^2)^\perp(\mu_\pi^2)^\perp+6\Big((\mu_\pi^2)^\perp\Big)^2\Big)\right.\nonumber\\
&\left.+\epsilon_{3/2}\Big(17\Big((\mu_G^2)^\perp\Big)^2+67(\mu_G^2)^\perp(\mu_\pi^2)^\perp+66\Big((\mu_\pi^2)^\perp\Big)^2\Big)\right]\  
\approx  14.71 \, {\rm GeV}^5  \,  , 
\end{align}
where we obtained the numerical estimate by using the above defined inputs. Taking the appropriate root gives $\sqrt[\leftroot{-2}\uproot{2}5]{X_{\rm{IC}}^{5}}\approx 1.7$ GeV.

 \begin{table}[t]
    \centering
    \begin{tabular}{|l||ll|}
    \multicolumn{3}{c}{\textbf{Input values}}\\
    \hline
        $m_b^{kin}$ &4.573 {\rm GeV} &\cite{Bordone:2021oof}\\
        $\bar{m}_c(2\ \text{GeV})$ & 1.092 {\rm GeV}&\cite{Bordone:2021oof}\\
    $\epsilon_{1/2}$&0.390 {\rm GeV}&\cite{Heinonen:2014dxa}\\
        $\epsilon_{3/2}$&0.476 {\rm GeV}&\cite{Heinonen:2014dxa}\\
        $(\mu_\pi^2)^\perp$&0.477 {\rm GeV}$^2$&\cite{Bordone:2021oof}\\
        $(\mu_G^2)^\perp$ &0.306 {\rm GeV}$^2$&\cite{Bordone:2021oof}\\
        \hline
    \end{tabular} \hfill 
    \begin{tabular}{|l||l|}
     \multicolumn{2}{c}{\textbf{LLSA approximation}}\\
        \hline
        $\mu_3$&0.996\\
        $\mu_G^2$&0.290 {\rm GeV}$^2$\\
        $\tilde{\rho}_D^3$ &0.205 {\rm GeV}$^3$\\
        \hline
        $\tilde{r}_E^4$ &0.098 {\rm GeV}$^4$\\
        $r_G^4$ &0.16 {\rm GeV}$^4$\\
        $\tilde{s}_E^4$ &-0.074 {\rm GeV}$^4$\\
        $s_B^4$&-0.14 {\rm GeV}$^4$\\
        $s_{qB}^4$ &-1.00 {\rm GeV}$^4$\\
        \hline
    \end{tabular} \hfill 
    \begin{tabular}{|l||l|}
    \multicolumn{2}{c}{\textbf{LLSA approximation}}\\
        \hline 
        $X_1^5$&0.049 {\rm GeV}$^5$\\
        $X_2^5$ &0.00  {\rm GeV}$^5$\\
        $X_3^5$ &0.094  {\rm GeV}$^5$\\
        $X_4^5$ &-0.41 {\rm GeV}$^5$\\
        $X_5^5$ &-0.039 {\rm GeV}$^5$\\
        $X_6^5$ &0.00  {\rm GeV}$^5$\\
        $X_7^5$&0.091 {\rm GeV}$^5$\\
        $X_8^5$&-0.0030 {\rm GeV}$^5$\\
        $X_9^5$&0.27 {\rm GeV}$^5$\\
        $X_{10}^5$ &0.025 {\rm GeV}$^5$\\
        \hline
    \end{tabular}
    \caption{The input values used for the numerical analysis are presented in the left table. The other two tables show the values for the RPI HQE parameters based on the LLSA approximation.}
    \label{tab:inputs}
\end{table}

 Finally, it is interesting to compare the estimate of the LLSA with the HQE parameters extracted from the $B\to X_c\ell \bar{\nu}$ spectra. Including only terms up to $
 1/m_b^3$, these fits yield $\rho_D^3 = (0.185\pm 0.031)$ GeV$^3$ \cite{Bordone:2021oof}. Using \eqref{LLSASR}, we may write\footnote{This expression holds up to $1/m_b^3$ terms. Introducing a commutator between the covariant derivatives (see \eqref{eq:mb4el}) also absorbs higher order terms into $\rho_D^3$ and thus would alter the LLSA expression.}
\begin{align}
    (\rho_D^3)^\perp &\equiv\frac{1}{2m_B}\langle \bar{b}_v\,(iD^\perp_\mu)(ivD)(iD^\perp_\nu)\,b_v\rangle g_\perp^{\mu\nu} \nonumber\\
    &= \frac{1}{3}\epsilon_{1/2} \big((\mu_\pi^2)^\perp-(\mu_G^2)^\perp\big) + \frac{1}{3}\epsilon_{3/2} \big(2(\mu_\pi^2)^\perp+(\mu_G^2)^\perp\big) = 0.22 \;{\rm GeV}^3\ .
\end{align}
This estimate is in good agreement with the value extracted from data. This strengthens our belief in the LLSA as a first estimate of the HQE parameters. In \cite{Gambino:2016jkc}, HQE elements up to $1/m_b^5$ in the full basis were extracted from moments of the lepton energy and $M_X$ spectra using the LLSA estimate to constrain their sizes. Allowing the fit to vary the HQE quantities, they found that most of the HQE elements changed very little with respect to their LLSA values and concluded that there was low sensitivity to the higher-power elements. In addition, in \cite{Bernlochner:2022ucr}, $r_E^4$ and $s_E^4$ were extracted from the $q^2$ moments, finding results consistent with zero. Within uncertainties they also agree with the values found in Table~\ref{tab:inputs}.

\subsection{\boldmath The total rate and the $q^2$ moments \unboldmath}
\begin{figure}[t!]
    \centering
    \includegraphics[width=0.48\textwidth]{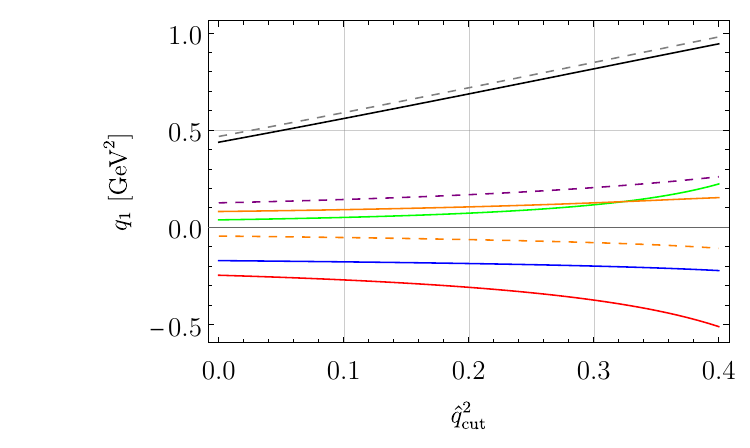}
    \includegraphics[width=0.48\textwidth]{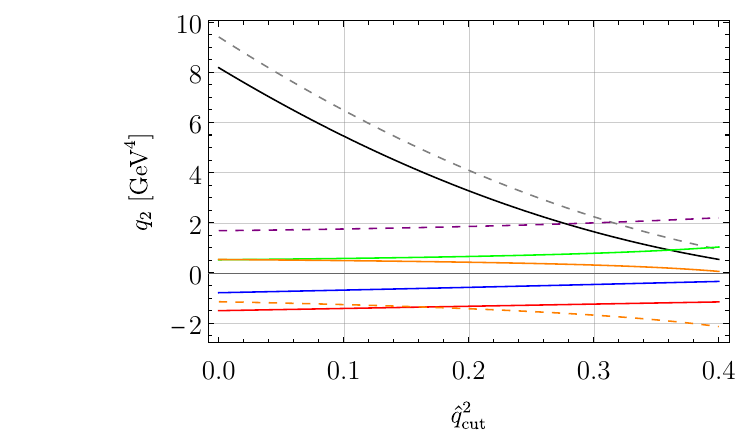}
    \includegraphics[width=0.48\textwidth]{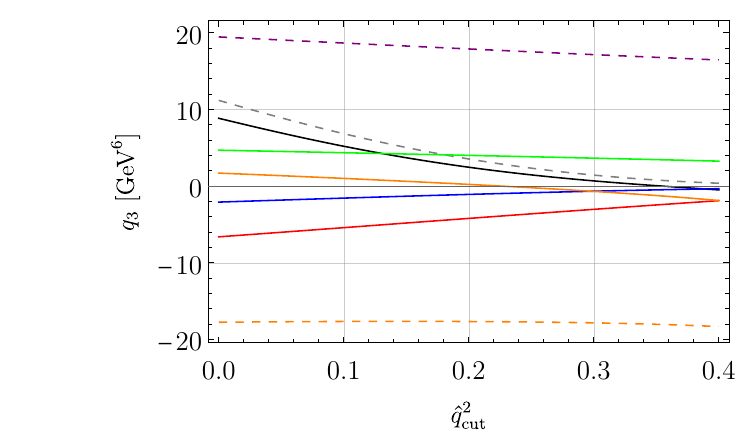}
    \includegraphics[width=0.48\textwidth]{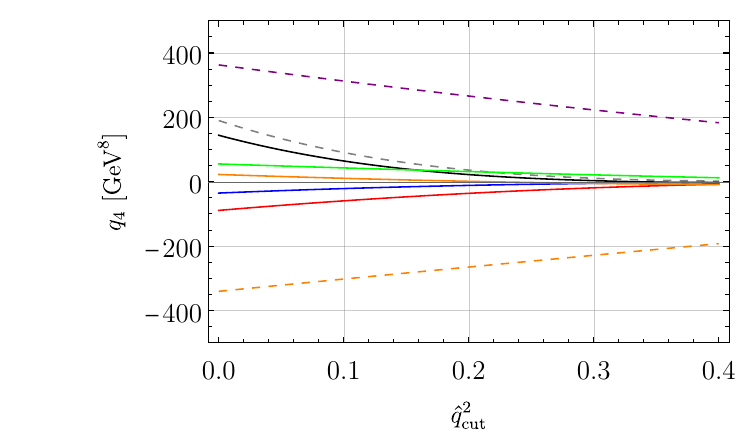}
    \includegraphics[width=\textwidth]{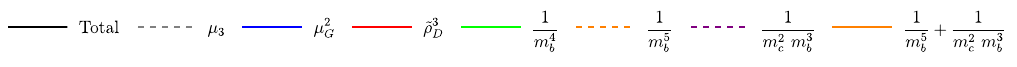}
    \caption{The dependence of  
    the first four centralized $q^2$ moments on the different orders in the $1/m_b$ expansion, as a function of the cut-off $\hat{q}^2_{\rm{cut}}$. The black dashed, blue solid, red solid, and green solid lines represent the contributions from $\mu_3$, $\mu_G^2$, $\tilde{\rho}_D^3$, and $1/m_b^4$ HQE parameters, respectively. The dashed orange, dashed purple, and solid orange lines represent the ``genuine'' $1/m_b^5$ contributions, IC contributions, and their summed contribution respectively. The black solid lines are the predictions for the centralized moments including contributions up to $1/m_b^5$. The total and $\mu_3$ results for $q_1$ have been divided by a factor of 10 to show the contributions at other orders more clearly.}
    \label{fig:qn_predictions}
\end{figure}

Using the LLSA estimates for the HQE elements, we can now consider the effects of the various contributions to observables. For simplicity, we focus here on RPI observables such as the total rate and the $q^2$ moments, including a 
cut on low values of $q^2$:
\begin{equation}
    \hat{q}^2_{\rm{cut}} \equiv \frac{{q}^2_{\rm{cut}} }{m_b^2} \ .
\end{equation}
In Fig.~\ref{fig:qn_predictions}, we show the various contributions fo the first four centralized $q^2$ moments $q_{1,2,3,4}$ as a function of $\hat{q}^2_{\rm{cut}}$. The black-dotted line is the leading term proportional to $\mu_3$, while
the black-solid line denotes the total contribution of all terms of the HQE up to dimension-8 operators. 
The colored lines show the individual contributions at each order in the HQE, where we have displayed the contributions 
of IC and the ``genuine'' $1/m_b^5$ pieces separately. As commented before, we do not show any uncertainties as we currently do not have a way to estimate the uncertainty associated to the LLSA. We find that, within the LLSA estimates, the IC and ``genuine'' $1/m_b^5$ terms are roughly equal in size, but contribute oppositely to the different $q^2$ moments. Most importantly, as discussed before, based on power-counting arguments the IC-parts would contribute at the same level as the $1/m_b^4$. We indeed observe that the IC contribution is large (in fact, larger than the $1/m_b^4$ terms in this estimate), but we also note that the other $1/m_b^5$ give large and opposite contributions. From this we conclude that, at least within the LLSA, only taking the IC-parts as part of a $1/m_b^4$ analysis could severely overestimate its effects,  and we thus recommend to consider these terms only in a combined determination up to $1/m_b^5$. We will return to this 
issue in Subsec.~\ref{sec:IC}.   

\begin{figure}[t!]
    \centering
    \includegraphics[width=0.48\textwidth]{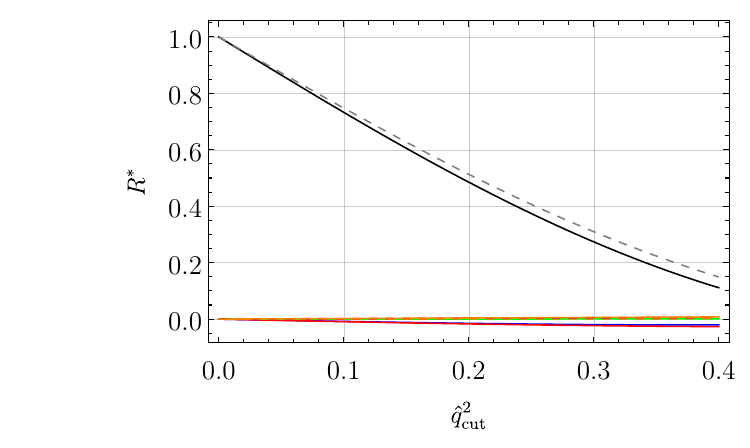} 
    \includegraphics[width=0.48\textwidth]{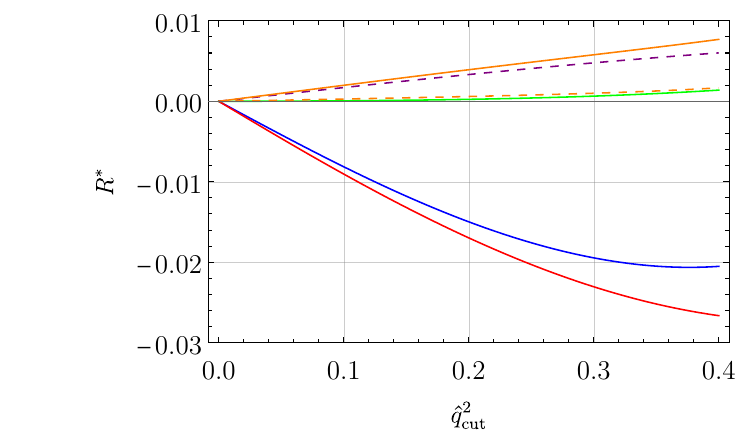}
    \includegraphics[width=\textwidth]{plotlegend.pdf}
    \caption{Similar as Fig.~\ref{fig:qn_predictions}, but now for the ratio $R^*$ and a zoom-in.}
    \label{fig:rstar}
\end{figure}

In Fig.~\ref{fig:rstar}, we show the ratio $R^*$ defined in \eqref{eq:rstar}. We observe that here the IC parts are larger than the $1/m_b^4$ terms and are not compensated by the genuine $1/m_b^5$ terms. So far, this ratio has only been measured with a lower cut on the lepton energy, which spoils its RPI behavior and introduces a dependence on the full set of HQE parameters. In trying to extract also the higher-order moments from data, it would be useful to have measurements of this observable given its different sensitivity compared to the $q^2$ moments. Finally, for the $|V_{cb}|$ extraction, the effect of the $1/m_b^5$ terms on the total rate is important. We list the full expression in Appendix~\ref{sec:ap_totalrate}. Numerically, employing the inputs of the previous section, we have\footnote{For $\tilde{\rho}_D^3$ we use here the estimate obtained in \cite{Fael:2018vsp}, using the determination in \cite{Gambino:2016jkc}: $\tilde{\rho}_D^3=0.127$ GeV$^3$.}
\begin{align}
    \frac{\Gamma(B\to X_c\ell\bar{\nu})}{\Gamma_0}=&\ 0.65\big|_{\mu_3}-0.22\big|_{\mu_G^2}-0.016\big|_{\tilde{\rho}_D^3}-0.00026\big|_{1/m_b^4}\nonumber\\
    &+0.0086\big|_{\rm{IC}}-0.0018\big|_{1/m_b^5}+\mathcal{O}(1/m_b^6)\ ,
\end{align}
where we indicate the effect of the different orders. We observe that the IC contribution and ``genuine'' $1/m_b^5$ contribute to the total rate with opposite sign. Note that this is different than for $R^*(\hat{q}^2_{\rm{cut}})$, as seen in Fig.\ \ref{fig:rstar} which also contains a term proportional to $\mu_G^2\tilde{\rho}_D^3/(\mu_3m_b^5)$ which the total rate does not.

\subsection{``Genuine'' \boldmath $1/m_b^5$ \unboldmath versus the IC contributions} \label{sec:IC}
Finally we discuss the observation that the ``genuine'' $1/m_b^5$ contribution almost cancels the 
IC pieces in the $q^2$ moments. 
We note that the Wilson coefficients $C$ of the operators at dimension-8 have for the total rate 
the generic form 
\begin{equation}
C = a_4 \rho^4 + a_3 \rho^3 + a_2 \rho^2 + a_1 \rho + a_0 + b_1 \frac{1}{\rho} + c_0 \ln \rho \ .
\end{equation}
The fact that the total rate has to vanish at $\rho = 1$ due to the vanishing phase space implies the 
relation 
\begin{equation} \label{Coeffrel} 
0 = a_4 + a_3 + a_2 + a_1 + a_0 + b_1  \quad \mbox{or} \quad  b_1 = - (a_4 + a_3 + a_2 + a_1 + a_0)\ .
\end{equation}
All the contributions proportional to the $a_i$ are attributed to the ``genuine'' $1/m_b^5$ while 
the logarithmic term and in particular the one with negative powers will be attributed to IC. 
The relation (\ref{Coeffrel}) links the two contributions, explaining at least part of the cancellation.

However, the details depend on the values of the HQE parameters, and for a detailed quantitative analysis 
we have to rely on the values obtained in LLSA shown in Table~\ref{tab:inputs}, in particular on the relative 
signs of the $X_i^5$ contributions predicted by LLSA. To obtain a general idea, 
we investigate this effect by taking a generic absolute value for the different matrix elements 
but vary their signs. In order to do this, we write  

\begin{align}
    q_n\supseteq \frac{m_b^{2n}}{\mu_3}\frac{1}{m_b^5}\Bigg(a^{m_b^5}_{n0}\frac{\mu_G^2\tilde{\rho}_D^3}{\mu_3}+\sum\limits_{i=1}^{10}a^{m_b^5}_{ni} X_i^5+\sum\limits_{j\in J}a^{\rm{IC}}_{nj}X_j^5\Bigg)\ ,\label{eq:cancellations}
\end{align}
where $J=\{2,3,5,7,8,9,10\}$, i.e.\ dimension-8 operators which contribute to the IC part (see \eqref{eq:XIC_def}). To obtain the expression for $R^*$, one sets $n=0$. To compare the genuine contributions to the IC contributions, we make the following assumptions. We assume that $X_i^5\sim\frac{\mu_G^2\tilde{\rho}^3_D}{\mu_3}\sim \Lambda_{\rm{QCD}}^5$. In addition, we assume that each $X_i^5$ contribution scales as $\Lambda_{\rm{QCD}}^5$, but we let the signs of these terms free. 

We can then write
\begin{align}
    q_n\supseteq \frac{m_b^{2n}}{\mu_3}\frac{1}{m_b^5}\Big(A_n^{m_b^5}+A_n^{\rm{IC}}\Big)\Lambda_{\rm{QCD}}^5\ ,
\end{align}
where
\begin{align}
A^{m_b^5}_n&\equiv a_{n0}^{m_b^5}+\sum\limits_{i=1}^{10}a_{ni}^{m_b^5}\cdot\text{sgn}\left(X_i^5\right)\ ,\\
    A^{\rm{IC}}_n&\equiv\sum\limits_{j\in J}a_{nj}^{\rm{IC}}\cdot\text{sgn}\left(X_j^5\right)\ ,
\end{align}
where these $A^k_n$ are functions of $\hat{q}^2_{\rm{cut}}$. 

\begin{figure}[h!]
    \centering
    \includegraphics[width=0.48\textwidth]{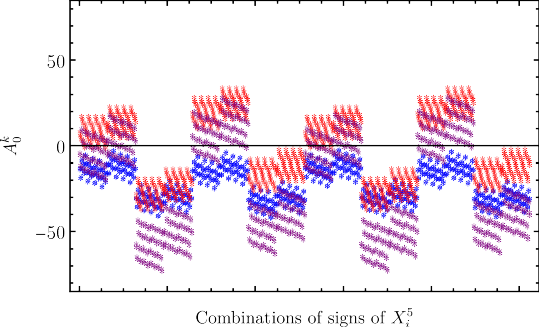}\includegraphics[width=0.3\textwidth]{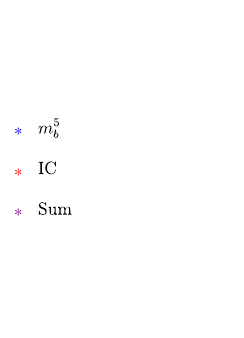}\\
    \vspace{0.3cm}
    \includegraphics[width=0.48\textwidth]{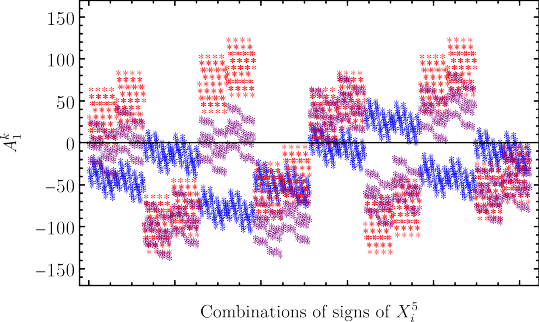}
    \includegraphics[width=0.48\textwidth]{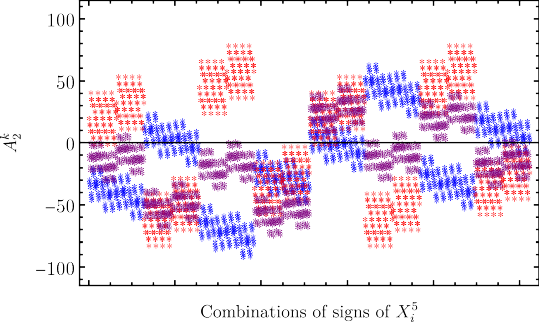}
    \includegraphics[width=0.48\textwidth]{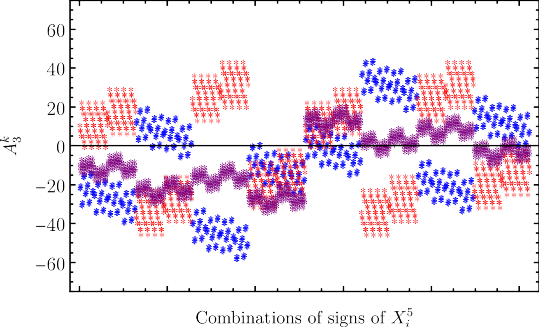}
    \includegraphics[width=0.48\textwidth]{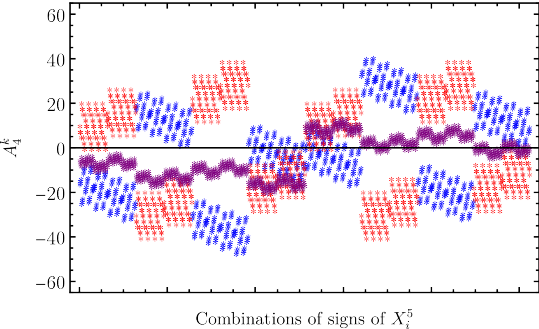}
    \caption{Relative contributions of the IC and the ``genuine" $1/m_b^5$ terms, and their sum to the centralized $q^2$ moments $q_n$ for $n=1,2,3,4$ and for the ratio $R^*$ ($n=0$). The x-axis enumerates the $2^{10}$ possible sign combinations of the HQE parameters $X_i^5$. }
    \label{fig:cancellations}
\end{figure}

In Fig.\ \ref{fig:cancellations}, we show the $A^{\rm IC}$, the $A^{m_b^5}$ and the sum of the two for all $2^{10}$ possible combinations of signs of the $X_i^5$ operators. From this, we see that the IC and genuine $1/m_b^5$ terms cancel each other to a large extent for most sign combinations, especially for the higher $q^2$ moments. Here, we show the results for $\hat{q}^2_{\rm{cut}}=0$, but we have explicitly checked that a similar behaviour applies to other $q^2$ cuts. For $R^*$, we used $\hat{q}^2_{\rm{cut}}=0.1$. Specifically, we find that if $\text{sgn}\big(X_3^5\big)=+1$, then the cancellation almost always occurs. This can be understood, because $X_3^5$ presents the dominant contribution of $X_{\rm{IC}}^{5}$ due to its large prefactor.  In light of our findings it seems advisable to also include all the $1/m_b^5$ terms and not only the IC contributions.
\clearpage
\section{Conclusion}\label{sec:conc}
We presented the complete tree-level contributions of the dimension-8 operators for the HQE of the differential rate 
for $B \to X_c \ell \bar{\nu}$. We used the standard form of the HQE where the bottom- and the charm-quark 
are integrated out at the same scale $\mu^2 \sim m_b m_c$. Consequently the Wilson coefficients of the HQE depend on the mass ration $\rho = m_c^2/m_b^2$. 

Starting at dimension six (corresponding to order $1/m_b^3$ in the HQE) an infrared sensitivity to the 
charm-quark mass arises as a $\log \rho$, while at higher orders also negative powers of 
$\rho$ appear. At tree level this happens first at order $1/m_b^5$, turning this into $1/(m_b^3 m_c^2)$. 
Adopting a power counting of the form $m_c^2 \sim \Lambda_{\rm QCD} m_b$ suggests, that such terms 
involving ``intrinsic charm'' should 
be counted as $1/m_b^4$, and hence could be parametrically larger than the ``genuine'' $1/m_b^5$ terms. 

We have derived ``trace formulae'' allowing us to compute any observable for $B \to X_c \ell \bar{\nu}$ 
up to dimension-8 terms, i.e. up to $1/m_b^5$ of the HQE. As known, the number of independent 
HQE parameters proliferates significantly. A minimum of ten HQE parameters enters at $1/m_b^5$, even 
if we take advantage of the reduced set of operators that enters in reparametrization invariant observables. For a phenomenological study of the effects of these higher orders, we thus have to estimate the values of the HQE parameters in some way, which we do by using the lowest-lying state saturation ansatz (LLSA). 

Using these values for the HQE parameters as a first estimate, we can make quantitative statements about the relative size 
of the previously unknown contributions of dimension-eight operators. However, it is currently challenging to assign an uncertainty to this estimate. We leave more detailed study of the higher-order corrections and the uncertainties of the LLSA to future work. 

We observe two interesting points.
\begin{enumerate}
    \item The size of the intrinsic-charm contributions is numerically of the expected (i.e. parametric) 
    size, however, the genuine $1/m_b^5$ terms contribute with the same magnitude.  
    \item The intrinsic charm contributions have the opposite sign as the ones from the genuine $1/m_b^5$
    terms, which leads in total to an unexpectedly small overall contribution of the dimension-eight operators.
\end{enumerate}
These statements at least holds in the LLSA, however, we have played with different scenarios, most of which 
exhibit a similar cancellation. Overall, there are indications that at least the $1/m_b^5$ contributions in the HQE are smaller than expected, 
which will allow us to reduce the theoretical uncertainty for the inclusive determination of $V_{cb}$ even further.

\clearpage
\section*{Acknowledgements}
This research has been supported by the Deutsche Forschungsgemeinschaft 
(DFG, German Research Foundation) under grant 396021762 - TRR 257.

 \appendix
\section{Conversion between different conventions\\ continued} \label{sec:ap_conversion}
\subsection{\boldmath $1/m_b^5$ definitions}
The $r_i$ are the full set of $1/m_b^5$ operators, including non-RPI operators, as defined in \cite{Mannel:2010wj}. For completeness, we list them here \cite{Mannel:2010wj}:
\begin{align}
\nonumber
2m_B r_1 &= \langle \bar{b}_v \,(i  D_\mu)\, (ivD)^3\, (i  D^\mu) \, b_v\rangle\ , \\
\nonumber 
2m_B r_2 &= \langle \bar{b}_v \,(i  D_\mu)\, (ivD)\, (i  D^\mu)\, (iD)^2 \, b_v\rangle\ , \\
\nonumber
2m_B r_3 &= \langle \bar{b}_v \,(i  D_\mu)\, (ivD)\,(i  D_\nu)\,( i
D^\mu)\,(i  D^\nu )\, b_v\rangle\ , \\  
\nonumber
2m_B r_4 &= \langle \bar{b}_v \,(i  D_\mu)\, (ivD)\, (iD)^2\,(i  D^\mu )\, b_v\rangle\ , \\  
\nonumber
2m_B r_5 &= \langle \bar{b}_v \,(iD)^2\,(ivD)\,  (iD)^2 \, b_v\rangle\ , \\  
\nonumber
2m_B r_6 &= \langle \bar{b}_v \,(i  D_\mu)\,(i  D_\nu)\, (ivD)\, (i
D^\nu)\,(i  D^\mu )\, b_v\rangle\ , \\  
\nonumber
2m_B r_7 &= \langle \bar{b}_v \,(i  D_\mu)\,(i  D_\nu)\, (ivD)\, (i
D^\mu)\,(i  D^\nu) \, b_v\rangle\ , \rule[-10pt]{0pt}{8pt}\\  
\nonumber
2m_B r_{8} &= \langle \bar{b}_v \,(i   D_\alpha) \, (ivD)^3\,( i   D_\beta)
\, (-i \sigma^{\alpha \beta })\,b_v\rangle\ , \\  
\nonumber
2m_B r_{9} &= \langle \bar{b}_v \,(i   D_\alpha) \, (ivD)\,( i   D_\beta)
\, (iD)^2 \,(-i \sigma^{\alpha \beta })\, b_v\rangle\ , \\  
\nonumber
2m_B r_{10} &= \langle \bar{b}_v \,(i   D_\mu)\, (ivD)\, (i
D^\mu)\, (i   D_\alpha) \,( i   D_\beta)  \,(-i \sigma^{\alpha \beta })\, b_v\rangle\ , \\
\nonumber
2m_B r_{11} &= \langle \bar{b}_v \,(i   D_\mu)\, (ivD)\, (i   D_\alpha)
\, (i   D^\mu)\,( i   D_\beta)  \,(-i \sigma^{\alpha \beta })\, b_v\rangle\ , \\  
\nonumber
2m_B r_{12} &= \langle \bar{b}_v \,(i   D_\alpha) \, (ivD)\, (i
D_\mu)\,( i   D_\beta) \, (i   D^\mu )\,(-i \sigma^{\alpha \beta })\, b_v\rangle\ , \\
\nonumber
2m_B r_{13} &= \langle \bar{b}_v \,(i   D_\mu)\, (ivD)\, (i   D_\alpha)
\,( i   D_\beta) \, (i   D^\mu )\,(-i \sigma^{\alpha \beta })\, b_v\rangle\ , \\  
\nonumber
2m_B r_{14} &= \langle \bar{b}_v \,(i   D_\alpha) \, (ivD)\, (iD)^2\,( i   D_\beta)  \,(-i \sigma^{\alpha \beta })\, b_v\rangle\ , \\
\nonumber
2m_B r_{15} &= \langle \bar{b}_v \,(i   D_\alpha) \,( i   D_\beta) \, (ivD)\, (iD)^2 \,(-i \sigma^{\alpha \beta })\, b_v\rangle\ , \\  
\nonumber
2m_B r_{16} &= \langle \bar{b}_v \,(i   D_\mu)\,( i   D_\alpha) \, (ivD)\,( i   D_\beta) \, (i   D^\mu )\,(-i \sigma^{\alpha \beta })\, b_v\rangle\ , \\  
\nonumber
2m_B r_{17} &= \langle \bar{b}_v \,(i   D_\alpha) \,( i   D_\mu)\, (ivD)\, (i   D^\mu)\,( i   D_\beta)  \,(-i \sigma^{\alpha \beta })\, b_v\rangle\ , \\  
2m_B r_{18} &= \langle \bar{b}_v \,(i   D_\mu)\, (i   D_\alpha) \, (ivD)\, (i   D^\mu)\,( i   D_\beta)  \,(-i \sigma^{\alpha \beta })\, b_v\rangle \ .\label{ri_definitions}
\end{align}
The RPI operators $X_i^5$ we derived in this work and defined in (\ref{eq:Xidef4},\ref{eq:Xidef10}) can then be written as
\begin{align}
    X_1^5 &= r_1\ , \nonumber\\
    X_2^5 &= 2r_6-2r_7\ ,\nonumber \\
    X_3^5 &= -r_2+2r_3-r_4+r_6-r_7\ , \nonumber\\
    X_4^5 &= 2r_2+4r_3+2r_4-2r_5-2r_6-2r_7\ ,\nonumber \\
    X_5^5 &= r_8\ , \nonumber\\
    X_6^5 &= r_{16}+r_{17}-2r_{18}\ ,\nonumber \\
    X_7^5 &= r_{11}-r_{12}-r_{13}+r_{14}+r_{16}-r_{18}\ ,\nonumber \\
    X_8^5 &= r_{11}+r_{12}-r_{13}+r_{16}-r_{18}-r_9\ ,\nonumber \\
    X_9^5 &=4r_{10}-2r_{15} \ ,\nonumber \\
    X_{10}^5 &= 2r_{10}+2r_{13}-2r_{15}\ . 
\end{align}

\subsection{\boldmath Differences at $1/m_b^4$}
In the following, we expand the discussion on the conversion between different HQE parameter definitions in \cite{Fael:2018vsp}. The covariant derivative can be split into a spatial and a time derivative via 
\begin{equation}\label{eq:perpbasis}
i D_\mu  = v_\mu\, ivD + iD_\mu^\perp \ .
\end{equation}
The HQE parameters can be defined with either the full covariant derivatives or with $iD^\perp$ as in \cite{Mannel:2010wj, Gambino:2016jkc}, which we will refer to as the ``perp''-basis. \\
Beside the RPI parameters in \eqref{eq:mb4el}, we list the non-RPI parameters up to $1/m_b^4$ \cite{Fael:2018vsp}:
\begin{align}
    2m_B\rho_{LS}^3&=\frac{1}{2}\langle \bar{b}_v\,\{(iD_\alpha),\,[(ivD),\,(iD_\beta)]\}\,(-i\sigma^{\alpha\beta})\,b_v\rangle\ ,\nonumber\\
    2m_B\delta\rho_D^4&=\frac{1}{2}\langle\bar{b}_v\,[(iD_\mu),\,[(iD)^2,\,(iD^\mu)]]\,b_v\rangle\ ,\nonumber\\
    2m_B\delta\rho_{LS}^4&=\frac{1}{2}\langle\bar{b}_v\,\{(iD_\alpha),\,[(iD)^2,\,(iD_\beta)]\}\,(-i\sigma^{\alpha\beta})\,b_v\rangle\ ,\nonumber\\
    2m_B\delta_{G1}^4&=\langle \bar{b}_v\,\left((iD)^2\right)^2\,b_v\rangle\ ,\nonumber\\
    2m_B\delta_{G2}^4&=\langle \bar{b}_v\,\{(iD)^2,\,(iD_\alpha)\,(iD_\beta)\}\,(-i\sigma^{\alpha\beta})\,b_v\rangle\ .
\end{align}
For completeness, it is interesting to consider the differences between these two bases up to $1/m_b^5$. We find 
\begin{align}
      (\mu_\pi^2)^\perp=&\ 2 m_b^2(1 - \mu_3) + \mu_G^2+\frac{1}{4m_b^2}\left[-\frac{r_G^4}{2}-s_B^4+\delta_{G1}^4+\delta_{G2}^4\right]\nonumber\\
      &+\frac{1}{4m_b^3}\big[-r_9-r_{10}+r_{11}+r_{12}-2 r_{13}+r_{15}+r_{16}-r_{18}\big]\ ,\nonumber\\
      (\mu_G^2)^\perp=&\ \mu_G^2+\frac{1}{m_b}\big[\tilde{\rho}_D^3+\rho_{LS}^3\big]-\frac{1}{2m_b^2}\delta\rho_D^4\ ,\nonumber\\
      (\rho_D^3)^\perp=&\ \tilde{\rho}_D^3-\frac{1}{2m_b}\delta\rho_D^4\ ,\nonumber\\
      (\rho_{LS}^3)^\perp=&\ \rho_{LS}^3-\frac{1}{2m_b}\big[r_E^4+s_E^4\big]+\frac{1}{2m_b^2}\Big[-r_2+r_3-r_4+\frac{r_5}{2}+\frac{r_6}{2}-\frac{r_7}{2}-r_9-r_{10}\nonumber\\
      &+r_{11}+r_{12}-r_{13}-r_{14}+r_{15}+\frac{r_{16}}{2}+\frac{r_{17}}{2}-r_{18}\Big]\ .
\end{align}
Note that interestingly, $(\mu_G^2)^\perp$ does not get additional $r_i$ contributions when writing $\rho^3_{LS}$ in the full derivative basis. We also note that in \eqref{eq:RPIcompletion_rEsE}, we defined $\tilde{r}_E^4$ and $\tilde{s}_E^4$ operators, where
\begin{align}
    \tilde{r}_E^4&=r_E^4+\frac{1}{m_b}\big[r_2-r_4\big]\ ,\nonumber\\
    \tilde{s}_E^4&=s_E^4+\frac{1}{m_b}\big[r_9-r_{14}\big]\ .
\end{align}

The relations between the dimension-7 parameters $m_i$ (first introduced in \cite{Mannel:2010wj}) and the RPI parameters have been presented in \cite{Fael:2018vsp}. Here, we extend these relations by adding the dimension-8 corrections $r_i$ via\footnote{This equation corrects a typo in \cite{Fael:2018vsp} for the $m_1$ coefficient.}
\begin{align}
    m_1&=\frac{1}{3}\left(r_E^4+\frac{r_G^4}{2}+2\delta\rho_{D}^4+3\delta_{G1}^4\right)+\frac{1}{3m_b}\big[2r_2+r_5+2r_{10}+r_{15}\big]\ ,\nonumber\\
    m_2&=-r_E^4\ ,\nonumber\\
    m_3&=-2r_E^4+r_G^4\ ,\nonumber\\
    m_4&=2r_E^4-2r_G^4-2\delta\rho_{D}^4+\frac{2}{m_b}\big[-r_2+r_5-r_{10}+r_{15}\big]\ ,\nonumber\\
    m_5&=-s_E^4\ ,\nonumber\\
    m_6&=-s_B^4+s_E^4-\frac{1}{m_b} \big[-r_2+2 r_3-r_4+r_6-r_7-r_{10}+r_{11}+r_{12}-r_{14}+r_{17}-r_{18}\big]\ ,\nonumber\\
   m_7&=2\delta\rho_{LS}^4+2s_E^4+\frac{s_{qB}^4}{2}-\frac{2}{m_b} \big[ r_1-2 r_4+ r_7+ r_8- r_9+ r_{12}- r_{13}-2 r_{14}+ r_{18}\big]\ ,\nonumber\\
   m_8&=4\delta_{G2}^4+\frac{4}{m_b} \big[- r_1+ r_2+ r_4- r_5+ r_6- r_7- r_8+ r_9+ r_{14}+ r_{16}+ r_{17}-2 r_{18}\big]\ ,\nonumber\\
   m_9&=-2s_B^4+2s_E^4+\frac{s_{qB}^4}{2}-\frac{1}{m_b}\big[r_1-3 r_2+2 r_3-3 r_4+r_5+r_6+r_7+r_8-r_9-2 r_{10}\nonumber\\
   &\hspace{5.5cm}+2 r_{12}-3 r_{14}-r_{16}+r_{17}+2 r_{18}\big]\ .
\end{align}
The expressions in (\ref{IC_gamma5},\ref{IC_gamma}) for the calculation of the IC contribution result in five operators $\tilde{f}_i$ describing the intrinsic charm, as defined in \cite{Bigi:2009ym}. The conversions between the $\tilde{f}_i$ from \cite{Bigi:2009ym} and $r_i$ and our RPI operators at $1/m_b^5$ are given by
\begin{align}
    \tilde{f}_1&=-2r_2-4r_3-2r_4+2r_5+4r_7= -X_2^5-X_4^5\ ,\nonumber\\
    \tilde{f}_2&=-2r_2-4r_3-2r_4+2r_5+2r_6+2r_7= -X_4^5\ ,\nonumber\\
    \tilde{f}_3&=-2r_2+4r_3-2r_4+2r_6-2r_7= 2X_3^5\ ,\nonumber\\
    \tilde{f}_4&=-2r_9-4r_{10}+4r_{12}-2r_{14}+2r_{15}= -2X_7^5+2X_8^5-X_9^5\ ,\nonumber\\
    \tilde{f}_5&=2r_{10}+2r_{11}-2r_{12}-4r_{13}+r_{14}+2r_{16}-2r_{18}= 2X_7^5+X_9^5-X_{10}^5\ .
\end{align}

The operator describing the $1/(m_b^3m_c^2)$ intrinsic charm contribution can be written in terms of $\tilde{f}_i$ and our dimension-8 RPI operators as
\begin{align}
    X_{\rm{IC}}^{5}&=-24X_2^5+78X_3^5-12X_4^5+10X_7^5-20X_8^5+5X_9^5+5X_{10}^5\nonumber\\
    &= 24\tilde{f}_1-12\tilde{f}_2+39\tilde{f}_3-10\tilde{f}_4-5\tilde{f}_5\label{eq:fiIC} \ .
\end{align}
The combination of operators $\tilde{f}_i$ in \eqref{eq:fiIC} agrees with the IC contribution for\newline $\Gamma(B\to X_c\ell\bar{\nu})$ presented in \cite{Bigi:2009ym}.

\section{\boldmath Determination of $1/m_Q^5$ RPI operators\unboldmath}\label{sec:ap_rpiderivation}
Starting from \eqref{eq:rpirel}, we require the tensor decomposition of $C_{\mu\alpha\beta\nu}^{(4)}$ and $C_{\mu\alpha\beta\delta\nu}^{(5)}$. The tensor decomposition of $C_{\mu\alpha\beta\nu}^{(4)}$ (dropping single $\gamma$ matrices and $\slashed v$, considering only hermitian and parity-even operators\footnote{We refer to \cite{Mannel:2018mqv} for the argumentation why single $\gamma$ matrices and $\slashed v$ can be dropped.}) is given by \cite{Mannel:2018mqv}
\begin{align}
    C^{(4)}_{\mu\alpha\beta\nu}(v)=&\ y_1^{(4)}g_{\mu\nu}g_{\alpha\beta}+y_2^{(4)}g_{\mu\alpha}g_{\nu\beta}+y_3^{(4)}g_{\mu\beta}g_{\nu\alpha}\nonumber\\
    &+z_1^{(4)}v_\alpha v_\beta g_{\mu\nu}+z_2^{(4)}v_\mu v_\nu g_{\alpha\beta}+z_3^{(4)}[v_\mu v_\alpha g_{\beta\nu}+v_\nu v_\beta g_{\mu\alpha}]+z_4^{(4)}[v_\mu v_\beta g_{\alpha\nu}+v_\nu v_\alpha g_{\beta\mu}]\nonumber\\
    &+w^{(4)}v_\mu v_\alpha v_\beta v_\nu\nonumber\\
    &+\alpha_1^{(4)}(-i\sigma_{\mu\nu})g_{\alpha\beta}+\alpha_2^{(4)}(-i\sigma_{\alpha\beta})g_{\mu\nu}+\alpha_3^{(4)}[(-i\sigma_{\mu\alpha})g_{\beta\nu}+(-i\sigma_{\beta\nu})g_{\mu\alpha}]\nonumber\\
    &\hspace{0.5cm}+\alpha_4^{(4)}[(-i\sigma_{\mu\beta})g_{\alpha\nu}+(-i\sigma_{\alpha\nu})g_{\mu\beta}]\nonumber\\
    &+\beta_1^{(4)}(-i\sigma_{\mu\nu})v_\alpha v_\beta+\beta_2^{(4)}(-i\sigma_{\alpha\beta})v_\mu v_\nu+\beta_3^{(4)}[(-i\sigma_{\mu\alpha})v_\beta v_\nu+(-i\sigma_{\beta\nu})v_\mu v_\alpha]\nonumber\\
    &\hspace{0.5cm}+\beta_4^{(4)}[(-i\sigma_{\mu\beta})v_\nu v_\alpha+(-i\sigma_{\alpha\nu})v_\mu v_\beta]\ .
\end{align}
For simplicity, we consider the spin-independent and spin-dependent (those with $\sigma$ terms) separately. 

\subsection{Spin-independent operators}
The tensor decomposition of the spin-independent (SI) terms of $C_{\mu\alpha\beta\delta\nu}^{(5)}$ (dropping single $\gamma$ matrices and $\slashed v$, considering only hermitian and parity-even operators) is
\begin{align}
    C_{\mu\alpha\beta\delta\nu}^{(5,\ \rm{SI})}\ (v)=&\ y_1^{(5)}[g_{\mu\alpha}g_{\beta\delta}v_\nu+g_{\nu\delta}g_{\beta\alpha}v_\mu]+y_2^{(5)}[g_{\mu\alpha}g_{\beta\nu}v_\delta+g_{\nu\delta}g_{\beta\mu}v_{\alpha}]\nonumber\\
    &\hspace{0.5cm}+y_3^{(5)}g_{\mu\alpha}g_{\delta\nu}v_\beta+y_4^{(5)}[g_{\mu\beta}g_{\alpha\delta}v_\nu+g_{\nu\beta}g_{\delta\alpha}v_\mu]\nonumber\\
    &\hspace{0.5cm}+y_5^{(5)}[g_{\mu\beta}g_{\alpha\nu}v_\delta+g_{\nu\beta}g_{\delta\mu}v_\alpha]+y_6^{(5)}[g_{\mu\delta}g_{\alpha\beta}v_\nu+g_{\nu\alpha}g_{\delta\beta}v_\mu]\nonumber\\
    &\hspace{0.5cm}+y_7^{(5)}g_{\mu\delta}g_{\alpha\nu}v_\beta+y_8^{(5)}[g_{\mu\nu}g_{\alpha\beta}v_\delta+g_{\nu\mu}g_{\delta\beta}v_\alpha]+y_9^{(5)}g_{\mu\nu}g_{\alpha\delta}v_\beta\nonumber\\
    &+z_1^{(5)}[g_{\mu\alpha}v_\beta v_\delta v_\nu+g_{\delta\nu}v_\beta v_\alpha v_\mu]+z_2^{(5)}[g_{\mu\beta}v_\alpha v_\delta v_\nu+g_{\nu\beta}v_\delta v_\alpha v_\mu]\nonumber\\
    &\hspace{0.5cm}+z_3^{(5)}[g_{\mu\delta}v_\alpha v_\beta v_\nu+g_{\nu\alpha}v_\delta v_\beta v_\mu]+z_4^{(5)}g_{\mu\nu}v_\alpha v_\beta v_\delta\nonumber\\
    &\hspace{0.5cm}+z_5^{(5)}[g_{\alpha\beta}v_\mu v_\delta v_\nu+g_{\delta\beta}v_\nu v_\alpha v_\mu]+z_6^{(5)}g_{\alpha\delta}v_\mu v_\beta v_\nu\nonumber\\
    &+w^{(5)}v_\mu v_\alpha v_\beta v_\delta v_\nu\ .
\end{align}
Then, \eqref{eq:rpirel} gives the following equations of motion:
\begin{align}
    z_1^{(4)}&=m_Q(y_5^{(5)}+y_7^{(5)}+2y_8^{(5)}+y_9^{(5)})\ ,\nonumber\\
    z_2^{(4)}&=m_Q(2y_1^{(5)}+y_4^{(5)}+y_6^{(5)}+y_8^{(5)})\ ,\nonumber\\
    z_3^{(4)}&=m_Q(2y_1^{(5)}+y_2^{(5)}+y_4^{(5)}+y_6^{(5)})=m_Q(y_2^{(5)}+2y_3^{(5)}+y_5^{(5)}+y_8^{(5)})\ ,\nonumber\\
    z_4^{(4)}&=m_Q(2y_2^{(5)}+y_5^{(5)}+y_7^{(5)}+y_9^{(5)})=m_Q(2y_4^{(5)}+y_5^{(5)}+2y_6^{(5)})\ ,\nonumber\\
    w^{(4)}&=m_Q(2z_1^{(5)}+z_2^{(5)}+z_3^{(5)}+z_4^{(5)})=m_Q(z_2^{(5)}+z_3^{(5)}+2z_5^{(5)}+z_6^{(5)}) \ .\label{eq:z_formules}
\end{align}
From \cite{Mannel:2018mqv}, we take the equations of motion
\begin{align}
x_3^{(3)}&=2m_Q(z_2^{(4)}+z_4^{(4)})=m_Q(2z_3^{(4)}+z_1^{(4)}+z_4^{(4)})\ .\label{eq:rpi4relations}
\end{align}
The two contributions to the spin-independent part of $\mathcal{R}^{(5)}$ are
 \begin{align}
    \mathcal{R}_1^{(5,\ \rm{SI})}&= \sum\limits_{i=1}^{9}y_i^{(5)}O_i^{(5)}\ ,\\
    \mathcal{R}_2^{(5,\ \rm{SI})}&= \sum\limits_{i=1}^{6}z_i^{(5)}P_i^{(5)}\ ,
\end{align}
with the basis operators
\begin{align}
    O_1^{(5)}&=\bar{Q}_v \left\{\left((iD)^2\right)^2,(iv D)\right\}Q_v\ ,\nonumber\\
    O_2^{(5)}&=\bar{Q}_v \left\{(iD)^2,(iD_\mu)(iv D)(iD^\mu)\right\}Q_v\ ,\nonumber\\
    O_3^{(5)}&=\bar{Q}_v\, (iD)^2(ivD)(iD)^2\,Q_v\ ,\nonumber\\
    O_4^{(5)}&=\bar{Q}_v \left\{(iD_\mu)(iD_\nu)(iD^\mu)(iD^\nu),(iv D)\right\}Q_v\ ,\nonumber\\
    O_5^{(5)}&=\bar{Q}_v \left\{(iD_\mu)(iD_\nu),(iD^\mu)(iv D)(iD^\nu)\right\}Q_v\ ,\nonumber\\
    O_6^{(5)}&=\bar{Q}_v \left\{(iD_\mu)(iD)^2(iD^\mu),(iv D)\right\}Q_v\ ,\nonumber\\
    O_7^{(5)}&=\bar{Q}_v\, (iD_\mu)(iD_\nu)(ivD)(iD^\mu)(iD^\nu)\,Q_v\ ,\nonumber\\
    O_8^{(5)}&=\bar{Q}_v \left((iD_\mu)(iD)^2(ivD)(iD^\mu)+(iD_\mu)(ivD)(iD)^2(iD^\mu)\right)Q_v\ ,\nonumber\\
    O_9^{(5)}&=\bar{Q}_v\, (iD_\mu)(iD_\nu)(ivD)(iD^\nu)(iD^\mu)\,Q_v\ ,\nonumber\\
    P_1^{(5)}&=\bar{Q}_v \left\{(iD)^2,(iv D)^3\right\}Q_v\ ,\nonumber\\
    P_2^{(5)}&=\bar{Q}_v \left\{(iD_\mu)(ivD)(iD^\mu),(iv D)^2\right\}Q_v\ ,\nonumber\\
    P_3^{(5)}&=\bar{Q}_v \left\{(iD_\mu)(ivD)^2(iD^\mu),(iv D)\right\}Q_v\ ,\nonumber\\
    P_4^{(5)}&=\bar{Q}_v\, (iD_\mu)(ivD)^3(iD^\mu)\,Q_v\ ,\nonumber\\
    P_5^{(5)}&=\bar{Q}_v \left\{(ivD)(iD)^2(ivD),(iv D)\right\}Q_v\ ,\nonumber\\
    P_6^{(5)}&=\bar{Q}_v\, (ivD)(iD_\mu)(ivD)(iD^\mu)(ivD)\,Q_v\ .
\end{align}
Solving (\ref{eq:z_formules},\ref{eq:rpi4relations}) for the coefficients $y_i^{(5)}$ gives
\begin{align}
    \mathcal{R}_1^{(5,\ \rm{SI})}=&\ \frac{x_3^{(3)}}{4m_Q^2}\left[O^{(5)}_1+O_3^{(5)}\right]+\frac{z_4^{(4)}}{2m_Q}\left[-\frac{3}{2}O_1^{(5)}+O_2^{(5)}-O_3^{(5)}+O_4^{(5)}\right]\nonumber\\
    &+\frac{z_1^{(4)}}{2m_Q}\left[\frac{1}{2}O_1^{(5)}-O_2^{(5)}-O_3^{(5)}-O_4^{(5)}+2O_5^{(5)}\right]\nonumber\\
    &+y_6^{(5)}\left[-O_4^{(5)}+O_6^{(5)}\right]\nonumber\\
    &+y_7^{(5)}\left[-\frac{1}{4}O_1^{(5)}+\frac{1}{2}O_3^{(5)}+\frac{1}{2}O_4^{(5)}-O_5^{(5)}+O_7^{(5)}\right]\nonumber\\
    &+y_8^{(5)}\left[-O_1^{(5)}+O_2^{(5)}+O_4^{(5)}+O_8^{(5)}-2O_5^{(5)}\right]\nonumber\\
    &+y_9^{(5)}\left[-\frac{1}{4}O_1^{(5)}+\frac{1}{2}O_3^{(5)}+\frac{1}{2}O_4^{(5)}-O_5^{(5)}+O_9^{(5)}\right]\ ,\label{eq:R1SI_solved}
\end{align}
From \eqref{eq:R1SI_solved}, we can see that:
\begin{itemize}
    \item The operators proportional to $x_3^{(3)}$ and $z_{1,4}^{(5)}$ are part of ``RPI completions'' of lower order operators. As an example, the operators proportional to $x_3^{(3)}$ are part of the RPI completion of the $\bar{Q}_v\,(ivD)^3\, Q_v$ term at third order:
    \begin{align}
        \bar{Q}_v\,(ivD)^3\,Q_v\xrightarrow{\text{RPI compl.}}&\ \bar{Q}_v\left(ivD+\frac{1}{2m_Q}(iD)^2\right)^3 Q_v\nonumber\\
        &\supseteq \frac{1}{4m_Q^2}\left[O^{(5)}_1+O_3^{(5)}\right]\nonumber\\
        &=\frac{1}{4m_Q^2}\bar{Q}_v\Big((iD)^2(ivD)(iD)^2+\left((iD)^2\right)^2(ivD)\nonumber\\
        &\hspace{0.5cm}+(ivD)\left((iD)^2\right)^2\Big)Q_v\ . \label{rpicompl}
    \end{align}
    \item The operators proportional to $y_6^{(5)}$ consists of operators with an $ivD$ acting on a quark field and therefore will only contribute at higher order.
    \end{itemize}

 We are therefore left with three independent RPI operators, i.e. the operators proportional to $y_{7,8,9}^{(5)}$ (dropping higher order operators):
    \begin{align}
        \mathcal{K}^{(5,\ \rm{RPI})}_1&\equiv\frac{1}{2}O_3^{(5)}-O_5^{(5)}+O_7^{(5)}=2m_B(\frac{1}{2}r_2-2r_3+r_7)\ ,\nonumber\\
        \mathcal{K}^{(5,\ \rm{RPI})}_2&\equiv O_2^{(5)}-2O_5^{(5)}+O_8^{(5)}=2m_B(2r_2-4r_3+2r_4)\ ,\nonumber\\
        \mathcal{K}^{(5,\ \rm{RPI})}_3&\equiv\frac{1}{2}O_3^{(5)}-O_5^{(5)}+O_9^{(5)}=2m_B(-2r_3+\frac{1}{2}r_5+r_6)\ ,
    \end{align}
    where we have also included the conversion to the $r_i$ parameters for $Q=b$, given in \eqref{ri_definitions}.

Solving \eqref{eq:z_formules} for the coefficients $z_i^{(5)}$ gives
\begin{align}
    \mathcal{R}_2^{(5,\ \rm{SI})}=&\ \frac{w^{(4)}}{2m_Q}[P_1^{(5)}+P_5^{(5)}]+z_2^{(5)}[P_2^{(5)}-\frac{1}{2}P_1^{(5)}-\frac{1}{2}P_5^{(5)}]+z_3^{(5)}[P_3^{(5)}-\frac{1}{2}P_1^{(5)}-\frac{1}{2}P_5^{(5)}]\nonumber\\
    &+z_4^{(5)}[P_4^{(5)}-\frac{1}{2}P_1^{(5)}]+z_6^{(5)}[P_6^{(5)}-\frac{1}{2}P_5^{(5)}]\ \label{eq:R2solved}.
\end{align}
From \eqref{eq:R2solved}, we can see that:
\begin{itemize}
    \item The operators proportional to $w^{(4)}$ are part of the RPI completion of the $\bar{Q}_v\,(ivD)^4\,Q_v$ term at fourth order.
    \item The operators proportional to $z_{2,3,6}^{(5)}$ have an $ivD$ acting on a quark field and therefore only contribute at higher orders.
    \end{itemize}

We thus have one independent RPI operator (dropping higher order operators):
    \begin{align}
        \mathcal{K}^{(5,\ \rm{RPI})}_4&\equiv P_4^{(5)}=2m_Br_1\ .
    \end{align}
All in all, we find 4 spin-independent RPI operators at dimension-8.
\subsection{Spin-dependent operators}
The tensor decomposition of the spin-dependent ($\sigma$) terms of $C_{\mu\alpha\beta\delta\nu}^{(5)}$ (considering only hermitian and parity-even operators) is given by
\begin{align}
    C_{\mu\alpha\beta\delta\nu}^{(5,\ \sigma)}\ (v)=&\ \alpha_1^{(5)}[(-i\sigma_{\mu\alpha})g_{\beta\delta}v_\nu+(-i\sigma_{\delta\nu})g_{\alpha\beta}v_\mu]+\alpha_2^{(5)}[(-i\sigma_{\beta\delta})g_{\mu\alpha}v_\nu+(-i\sigma_{\alpha\beta})g_{\delta\nu}v_\mu]\nonumber\\
    &\hspace{0.5cm}+\alpha_3^{(5)}[(-i\sigma_{\mu\alpha})g_{\beta\nu}v_\delta+(-i\sigma_{\delta\nu})g_{\mu\beta}v_{\alpha}]+\alpha_4^{(5)}[(-i\sigma_{\beta\nu})g_{\mu\alpha}v_\delta+(-i\sigma_{\mu\beta})g_{\delta\nu}v_{\alpha}] \nonumber\\
    &\hspace{0.5cm}+\alpha_5^{(5)}[(-i\sigma_{\mu\alpha})g_{\delta\nu}v_\beta+(-i\sigma_{\delta\nu})g_{\alpha\mu}v_\beta]+\alpha_6^{(5)}[(-i\sigma_{\mu\beta})g_{\alpha\delta}v_\nu+(-i\sigma_{\beta\nu})g_{\alpha\delta}v_\mu] \nonumber\\
    &\hspace{0.5cm}+\alpha_7^{(5)}[(-i\sigma_{\alpha\delta})g_{\mu\beta}v_\nu+(-i\sigma_{\alpha\delta})g_{\beta\nu}v_\mu]+\alpha_8^{(5)}[(-i\sigma_{\mu\beta})g_{\alpha\nu}v_\delta+(-i\sigma_{\beta\nu})g_{\mu\delta}v_\alpha] \nonumber\\
    &\hspace{0.5cm}+\alpha_9^{(5)}[(-i\sigma_{\alpha\nu})g_{\mu\beta}v_\delta+(-i\sigma_{\mu\delta})g_{\beta\nu}v_\alpha]+\alpha_{10}^{(5)}[(-i\sigma_{\mu\delta})g_{\alpha\beta}v_\nu+(-i\sigma_{\alpha\nu})g_{\beta\delta}v_\mu] \nonumber\\
    &\hspace{0.5cm}+\alpha_{11}^{(5)}[(-i\sigma_{\alpha\beta})g_{\mu\delta}v_\nu+(-i\sigma_{\beta\delta})g_{\alpha\nu}v_\mu]+\alpha_{12}^{(5)}[(-i\sigma_{\mu\delta})g_{\alpha\nu}v_\beta+(-i\sigma_{\alpha\nu})g_{\delta\mu}v_\beta] \nonumber\\
    &\hspace{0.5cm}+\alpha_{13}^{(5)}[(-i\sigma_{\mu\nu})g_{\alpha\beta}v_\delta+(-i\sigma_{\mu\nu})g_{\delta\beta}v_\alpha]+\alpha_{14}^{(5)}[(-i\sigma_{\alpha\beta})g_{\mu\nu}v_\delta+(-i\sigma_{\beta\delta})g_{\nu\mu}v_\alpha] \nonumber\\
    &\hspace{0.5cm}+\alpha_{15}^{(5)}(-i\sigma_{\mu\nu})g_{\alpha\delta}v_\beta+\alpha_{16}^{(5)}(-i\sigma_{\alpha\delta})g_{\mu\nu}v_\beta \nonumber\\
    &+\beta_1^{(5)}[(-i\sigma_{\mu\alpha})v_\beta v_\delta v_\nu+(-i\sigma_{\delta\nu})v_\beta v_\alpha v_\mu]+\beta_2^{(5)}[(-i\sigma_{\mu\beta})v_\alpha v_\delta v_\nu+(-i\sigma_{\beta\nu})v_\delta v_\alpha v_\mu]\nonumber\\
    &\hspace{0.5cm}+\beta_3^{(5)}[(-i\sigma_{\mu\delta})v_\alpha v_\beta v_\nu+(-i\sigma_{\alpha\nu})v_\delta v_\beta v_\mu]+\beta_4^{(5)}(-i\sigma_{\mu\nu})v_\alpha v_\beta v_\delta\nonumber\\
    &\hspace{0.5cm}+\beta_5^{(5)}[(-i\sigma_{\alpha\beta})v_\mu v_\delta v_\nu+(-i\sigma_{\beta\delta})v_\nu v_\alpha v_\mu]+\beta_6^{(5)}(-i\sigma_{\alpha\delta})v_\mu v_\beta v_\nu\ .
\end{align}
Then \eqref{eq:rpirel} gives the following equations of motion
\begin{align}
\alpha^{(5)}_{6}-\alpha^{(5)}_{7}-\alpha^{(5)}_{9}&=0\ ,\nonumber\\
    \alpha^{(5)}_{4}+\alpha^{(5)}_{9}+\alpha^{(5)}_{13}&=0\ ,\nonumber\\
    \alpha^{(5)}_{8}+\alpha^{(5)}_{12}+\alpha^{(5)}_{15}&=0\ ,\nonumber\\
    \alpha^{(5)}_{8}-\alpha^{(5)}_{12}-\alpha^{(5)}_{16}&=0\ ,\nonumber\\
    \alpha^{(5)}_{9}+\alpha^{(5)}_{12}+2\alpha^{(5)}_{13}+\alpha^{(5)}_{15}&=\frac{\beta_1^{(4)}}{m_Q}\ ,\nonumber\\
    2\alpha^{(5)}_{2}+\alpha^{(5)}_{7}+\alpha^{(5)}_{11}+\alpha^{(5)}_{14}&=\frac{\beta_2^{(4)}}{m_Q}\ ,\nonumber\\
    \alpha^{(5)}_{3}+2\alpha^{(5)}_{5}+\alpha^{(5)}_{8}+\alpha^{(5)}_{14}&=\frac{\beta_3^{(4)}}{m_Q}\ ,\nonumber\\
    2\alpha^{(5)}_{1}+\alpha^{(5)}_{3}+\alpha^{(5)}_{6}+\alpha^{(5)}_{11}&=\frac{\beta_3^{(4)}}{m_Q}\ ,\nonumber\\
    \alpha^{(5)}_{6}+\alpha^{(5)}_{7}+\alpha^{(5)}_{8}+2\alpha^{(5)}_{10}&=\frac{\beta_4^{(4)}}{m_Q}\ ,\nonumber\\
    \beta_2^{(5)}+\beta_3^{(5)}+\beta_4^{(5)}&=0\ ,\nonumber\\
\beta_2^{(5)}-\beta_3^{(5)}-\beta_6^{(5)}&=0\ . \label{eq:beta_relations}
\end{align}
From \cite{Mannel:2018mqv}, we also take the equation of motion
\begin{align}
    \beta_1^{(4)}&=-\beta_4^{(4)}\ . \label{old_rpi_rel}
\end{align}
The two contributions to the spin-dependent part of $\mathcal{R}^{(5)}$ are
\begin{align}
    \mathcal{R}_1^{(5,\ \sigma)}&=\sum\limits_{i=1}^{6}\beta_i^{(5)}U_i^{(5)}\ ,\\
    \mathcal{R}_2^{(5,\ \sigma)}&=\sum\limits_{i=1}^{16}\alpha_i^{(5)}S_i^{(5)}\ ,
\end{align}
with operators
\begin{align}
    U_1^{(5)}&=\bar{Q}_v\left\{(\sigma\cdot G),(ivD)^3\right\} Q_v\ ,\nonumber\\
    U_2^{(5)}&=\bar{Q}_v\left\{(iD^\mu)(ivD)(iD^\nu),(ivD)^2\right\}(-i\sigma_{\mu\nu})\, Q_v\ ,\nonumber\\
    U_3^{(5)}&=\bar{Q}_v\left\{(iD^\mu)(ivD)^2(iD^\nu),(ivD)\right\}(-i\sigma_{\mu\nu})\, Q_v\ ,\nonumber\\
    U_4^{(5)}&=\bar{Q}_v\,(iD^\mu)(ivD)^3(iD^\nu)(-i\sigma_{\mu\nu})\,Q_v\ ,\nonumber\\
    U_5^{(5)}&=\bar{Q}_v\left\{(ivD)(\sigma\cdot G)(ivD),(ivD)\right\}Q_v\ ,\nonumber\\
    U_6^{(6)}&=\bar{Q}_v\,(ivD)(iD^\mu)(ivD)(iD^\nu)(ivD)(-i\sigma_{\mu\nu})\,Q_v\ ,\nonumber\\
    S_1^{(5)}&=\bar{Q}_v\left((\sigma\cdot G)(iD)^2(ivD)+(ivD)(iD)^2(\sigma\cdot G)\right)Q_v\ ,\nonumber\\
    S_2^{(5)}&=\bar{Q}_v\left((iD)^2(\sigma\cdot G)(ivD)+(ivD)(\sigma\cdot G)(iD)^2\right)Q_v\ ,\nonumber\\
    S_3^{(5)}&=\bar{Q}_v \left\{(\sigma\cdot G),(iD_\mu)(ivD)(iD^\mu)\right\}Q_v\ ,\nonumber\\
    S_4^{(5)}&=\bar{Q}_v\left\{(iD)^2,(iD^\alpha)(ivD)(iD^\beta)\right\}(-i\sigma_{\alpha\beta})\,Q_v\ ,\nonumber\\
    S_5^{(5)}&=\bar{Q}_v\left((\sigma\cdot G)(ivD)(iD)^2+(iD)^2(ivD)(\sigma\cdot G)\right)Q_v\ ,\nonumber\\
    S_6^{(5)}&=\bar{Q}_v\left((iD^\alpha)( iD_\mu)( iD^\beta)( iD^\mu)(ivD)+(ivD)(iD_\mu)( iD^\alpha)( iD^\mu)( iD^\beta)\right)(-i\sigma_{\alpha\beta})\,Q_v\ ,\nonumber\\
    S_7^{(5)}&=\bar{Q}_v\left((iD_\mu)( iD^\alpha)( iD^\mu)( iD^\beta)(ivD)+(ivD)(iD^\alpha)( iD_\mu)( iD^\beta)( iD^\mu)\right)(-i\sigma_{\alpha\beta})\,Q_v\ ,\nonumber\\
    S_8^{(5)}&=\bar{Q}_v\left((iD^\alpha)( iD_\mu)( iD^\beta)(ivD)(iD^\mu)+(iD_\mu)(ivD)(iD^\alpha)( iD^\mu)( iD^\beta)\right)(-i\sigma_{\alpha\beta})\,Q_v\ ,\nonumber\\
    S_9^{(5)}&=\bar{Q}_v\left((iD_\mu)( iD^\alpha)( iD^\mu)(ivD)(iD^\beta)+(iD^\alpha)(ivD)(iD_\mu)( iD^\beta)( iD^\mu)\right)(-i\sigma_{\alpha\beta})\,Q_v\nonumber \ ,\\
    S_{10}^{(5)}&=\bar{Q}_v\left((iD^\alpha)(iD)^2(iD^\beta)(ivD)+ (ivD)(iD^\alpha)(iD)^2(iD^\beta)\right)(-i\sigma_{\alpha\beta})\,Q_v\ ,\nonumber\\
    S_{11}^{(5)}&=\bar{Q}_v\left((iD_\mu)(\sigma\cdot G)(iD^\mu)(ivD)+(ivD)(iD_\mu)(\sigma\cdot G)(iD^\mu)\right)\,Q_v\ ,\nonumber\\
    S_{12}^{(5)}&=\bar{Q}_v\left((iD^\alpha)( iD_\mu)(ivD)(iD^\beta)( iD^\mu)+(iD_\mu)( iD^\alpha)(ivD)(iD^\mu)( iD^\beta)\right)(-i\sigma_{\alpha\beta})\,Q_v\ ,\nonumber\\
    S_{13}^{(5)}&=\bar{Q}_v\,(iD^\alpha)\left\{(iD)^2,(ivD)\right\}(iD^\beta)\,(-i\sigma_{\alpha\beta})\,Q_v\ ,\nonumber\\
    S_{14}^{(5)}&=\bar{Q}_v\,(iD_\mu)\left\{(\sigma\cdot G),(ivD)\right\}(iD^\mu)\,Q_v\ ,\nonumber\\
    S_{15}^{(5)}&=\bar{Q}_v\,(iD^\alpha)( iD_\mu)(ivD)(iD^\mu)( iD^\beta)\,(-i\sigma_{\alpha\beta})\,Q_v\ ,\nonumber\\
    S_{16}^{(5)}&=\bar{Q}_v\,(iD_\mu)( iD^\alpha)(ivD)(iD^\beta)( iD^\mu)\,(-i\sigma_{\alpha\beta})\,Q_v\ .
\end{align}
Solving the equations of motions in (\ref{eq:beta_relations},\ref{old_rpi_rel}) for $\alpha_i^{(5)}$ gives
\begin{align}
    \mathcal{R}_1^{(5,\ \sigma)}=&\ \frac{\beta_1^{(4)}}{2m_Q}\Big[S^{(5)}_{13}-S^{(5)}_{4}-S^{(5)}_{10}\Big]+\frac{\beta_2^{(4)}}{2m_Q}\Big[S^{(5)}_{2}\Big]+\frac{\beta_3^{(4)}}{2m_Q}\Big[S^{(5)}_{1}+S^{(5)}_{5}\Big]\nonumber\\
    &+\alpha_3^{(5)}\Big[S^{(5)}_{3}-\frac{1}{2}S^{(5)}_{1}-\frac{1}{2}S^{(5)}_{5}\Big]\nonumber\\
    &+\alpha_7^{(5)}\Big[S^{(5)}_{7}-\frac{1}{2}S^{(5)}_{1}-\frac{1}{2}S^{(5)}_{2}-S^{(5)}_{10}+S^{(5)}_{6}\Big]\nonumber\\
    &+\alpha_8^{(5)}\Big[S^{(5)}_{8}-\frac{1}{2}S^{(5)}_{5}-\frac{1}{2}S^{(5)}_{10}+\frac{1}{2}S^{(5)}_{13}-\frac{1}{2}S^{(5)}_{4}-S^{(5)}_{15}+S^{(5)}_{16}\Big]\nonumber\\
    &+\alpha_9^{(5)}\Big[S^{(5)}_{9}-\frac{1}{2}S^{(5)}_{1}-\frac{1}{2}S^{(5)}_{10}-\frac{1}{2}S^{(5)}_{13}+S^{(5)}_{6}-\frac{1}{2}S^{(5)}_{4}\Big]\nonumber\\
    &+\alpha_{11}^{(5)}\Big[S^{(5)}_{11}-\frac{1}{2}S^{(5)}_{1}-\frac{1}{2}S^{(5)}_{2}\Big]\nonumber\\
    &+\alpha_{12}^{(5)}\Big[S^{(5)}_{12}-S^{(5)}_{15}-S^{(5)}_{16}\Big]\nonumber\\
    &+\alpha_{14}^{(5)}\Big[S^{(5)}_{14}-\frac{1}{2}S^{(5)}_{2}-\frac{1}{2}S^{(5)}_{5}\Big]\ ,
\end{align}
The operators proportional to $\beta_i^{(4)}$ are parts of RPI completions. Dropping operators which have an $ivD$ acting on a quark field (since they only contribute at higher orders), we arrive at five independent RPI operators:

\begin{align}
    \mathcal{K}_5^{(5,\ \rm{RPI})}&\equiv S^{(5)}_{3}-\frac{1}{2}S^{(5)}_{5}=2m_B(2r_{10}-r_{15})\ , \nonumber\\
    \mathcal{K}_6^{(5,\ \rm{RPI})}&\equiv S^{(5)}_{9}-\frac{1}{2}S^{(5)}_{13}-\frac{1}{2}S^{(5)}_{4}=2m_B(2r_{12}-r_{14}-r_{9})\ , \nonumber\\
    \mathcal{K}_7^{(5,\ \rm{RPI})}&\equiv S^{(5)}_{14}-\frac{1}{2}S^{(5)}_{5}=2m_B(2r_{13}-r_{15})\ , \nonumber\\
    \mathcal{K}_8^{(5,\ \rm{RPI})}&\equiv S^{(5)}_{8}-\frac{1}{2}S^{(5)}_{5}+\frac{1}{2}S^{(5)}_{13}-\frac{1}{2}S^{(5)}_{4}-S^{(5)}_{15}+S^{(5)}_{16}\nonumber\\
    &=2m_B(2r_{11}-r_{15}+r_{14}-r_9-r_{17}+r_{16})\ ,\nonumber \\
    \mathcal{K}_{9}^{(5,\ \rm{RPI})}&\equiv S^{(5)}_{12}-S^{(5)}_{15}-S^{(5)}_{16}=2m_B(2r_{18}-r_{17}-r_{16})\ , 
\end{align}
Solving the relations in (\ref{eq:beta_relations},\ref{old_rpi_rel}) for $\beta_i^{(5)}$ gives
\begin{align}
    \mathcal{R}_2^{(5,\ \sigma)}=&\ \beta_1^{(5)}U_1^{(5)}+\beta_4^{(5)}\left[U_4^{(5)}-\frac{1}{2}U_2^{(5)}-\frac{1}{2}U_3^{(5)}\right]+\beta_5^{(5)}U_5^{(5)}\nonumber\\&+\beta_6^{(5)}\left[U_6^{(5)}+\frac{1}{2}U_2^{(5)}-\frac{1}{2}U_3^{(5)}\right]\ , \label{R2sigma_solved}
\end{align}
From \eqref{R2sigma_solved}, we can see that only the operator $U_4^{(5)}$ does not have an $ivD$ term acting directly on a quark field. The other terms thus only contribute at higher orders. Therefore, the only relevant RPI operator left is (dropping higher order operators)
    \begin{align}
        \mathcal{K}_{10}^{(5,\ \rm{RPI})}\equiv U_4^{(5)}=2m_Br_8\ .
    \end{align}
\subsection{RPI operators at dimension-8}
We find in total 10 independent RPI operators at dimension-8, of which 4 spin-independent and 6 spin-dependent. We defined the  $X_i^5$ in (\ref{eq:Xidef4},\ref{eq:Xidef10}) in Sec.\ \ref{sec:hqemb5} with commutators of covariant derivatives in order to allow for an interpretation in terms of gluon 
fields and gluon momenta. A basis transformation gives the RPI operators $X_i^5$ in terms of $\mathcal{K}_i^{(5,\ \rm{RPI})}$:
\begin{align}
    X_1^5&=\mathcal{K}^{(5,\ \rm{RPI})}_4\ ,\nonumber\\
    X_2^5    &=2\mathcal{K}^{(5,\ \rm{RPI})}_3-2\mathcal{K}^{(5,\ \rm{RPI})}_1\ ,\nonumber\\
    X_3^5&=\mathcal{K}^{(5,\ \rm{RPI})}_3-\mathcal{K}^{(5,\ \rm{RPI})}_1-\frac{1}{2}\mathcal{K}^{(5,\ \rm{RPI})}_2\ ,\nonumber\\
    X_4^5 &=\mathcal{K}^{(5,\ \rm{RPI})}_2-2\mathcal{K}^{(5,\ \rm{RPI})}_1-2\mathcal{K}^{(5,\ \rm{RPI})}_3\ ,\nonumber\\
    X_5^5&=\mathcal{K}^{(5,\ \rm{RPI})}_{10}\ ,\nonumber\\
    X_6^5&=-\mathcal{K}^{(5,\ \rm{RPI})}_9\ ,\nonumber\\
    X_7^5&=-\frac{1}{2}\mathcal{K}^{(5,\ \rm{RPI})}_6-\frac{1}{2}\mathcal{K}^{(5,\ \rm{RPI})}_7+\frac{1}{2}\mathcal{K}^{(5,\ \rm{RPI})}_8-\frac{1}{2}\mathcal{K}^{(5,\ \rm{RPI})}_9\ ,\nonumber\\
    X_8^5&=\frac{1}{2}\mathcal{K}^{(5,\ \rm{RPI})}_6-\frac{1}{2}\mathcal{K}^{(5,\ \rm{RPI})}_7+\frac{1}{2}\mathcal{K}^{(5,\ \rm{RPI})}_8-\frac{1}{2}\mathcal{K}^{(5,\ \rm{RPI})}_9\ ,\nonumber\\
    X_9^5&=2\mathcal{K}^{(5,\ \rm{RPI})}_5\ ,\nonumber\\
    X_{10}^5&=\mathcal{K}^{(5,\ \rm{RPI})}_5+\mathcal{K}^{(5,\ \rm{RPI})}_7\ .
\end{align}
\clearpage
\section{\boldmath Expressions for the trace formula and $q^2$ moments\unboldmath}\label{sec:ap_traceform}
The attached Mathematica notebook \texttt{Trace\_Formula\_v2.nb} contains expressions for trace formula $\mathcal{M}^{(n)}_{\mu_1...\mu_{n-3}}$, defined in \eqref{traceformulaslist}, for dimensions $n=k+3=3,...,8$ which can be used to calculate forward matrix elements through
   \begin{align}
        \langle B|\bar{b}_{v}\, (iD_{\mu_1})\,...\,(iD_{\mu_{k}})\,\Gamma_v^{\mu_1...\mu_{k}}\,b_{v}|B\rangle&=\text{Tr}\left[\mathcal{M}^{(k+3)}_{\mu_1...\mu_{3}}\Gamma_v^{\mu_1...\mu_{3}}\right]\ ,
    \end{align}
where $\Gamma_v^{\mu_1...\mu_{d-3}}$ is composed of Dirac matrices $\gamma^\mu$, metrics $g^{\mu\nu}$, and four-momenta $v^\mu$. The definitions of the matrix elements used in $\mathcal{M}^{(n)}_{\mu_1...\mu_{n-3}}$ are included in the notebook itself, and contain all (non-RPI) operators at each dimension. The expressions can therefore also be implemented to calculate non-RPI quantities, like lepton-energy moments.

We also include an ancillary  file \texttt{Q2\_moments\_with\_q2cut\_v2.nb} with expressions for 
\begin{align}
    \mathcal{Q}_n(\hat{q}^2_{\rm{cut}})&=\frac{1}{\Gamma_0}\int_{\hat{q}^2_{\rm{cut}}}^{(1-\sqrt{\rho})^2}\text{d}\hat{q}^2\ (\hat{q}^2)^n\frac{\text{d}\Gamma}{\text{d}\hat{q}^2}\ ,
\end{align}
for $n=0,1,2,3,4$ up to $1/m_b^5$. These can be used to determine the total rate, the ratio $R^*$, and the $q^2$ moments by re-expanding the following relations in $1/m_b$:
\begin{align}
    \Gamma= \Gamma_0 \mathcal{Q}_0(0)\ , \quad\quad R^*(\hat{q}^2_{\rm{cut}})=\frac{\mathcal{Q}_0(\hat{q}^2_{\rm{cut}})}{\mathcal{Q}_0(0)}\ , \quad \quad \langle (q^2)^n\rangle = m_b^{2n}\frac{\mathcal{Q}_n(\hat{q}^2_{\rm{cut}})}{\mathcal{Q}_0(\hat{q}^2_{\rm{cut}})} \ .
\end{align}
The definitions of the HQE parameters are included in the notebook itself. 

In the updated version, the expressions for the trace formula and $q^2$ moments in \texttt{Trace\_Formula\_v2.nb} and \texttt{Q2\_moments\_with\_q2cut\_v2.nb} have been updated, resulting in a difference proportional to $-X_8^5+\frac{1}{2}X_{10}^5$, as identified in \cite{Finauri:2025ost}.

\clearpage
\section{\boldmath Total rate up to $1/m_b^5$ \unboldmath}
\label{sec:ap_totalrate}
For completeness, we also present the total decay rate in terms of our matrix elements and 
$\rho = m_c^2/m_b^2$. 
The intrinsic charm contribution can be easily identified as the terms proportional to $1/\rho$. 
\begin{align}
    \frac{1}{\Gamma_0}&\Gamma(B\to X_c\ell\bar{\nu})= \mu_3\Big(1-8\rho+8\rho^3-\rho^4-12\rho^2\log{\rho}\Big)-\frac{2\mu_G^2}{m_b^2}\Big(1-4\rho+6\rho^2-4\rho^3+\rho^4\Big)\nonumber\\
    &+\frac{2\tilde{\rho}_D^3}{3m_b^3}\Big(17-16\rho-12\rho^2+16\rho^3-5\rho^4+12\log{\rho}\Big)
        -\frac{8\tilde{r}_E^4}{9m_b^4}\Big(2+9\rho^2-20\rho^3+9\rho^4+6\log{\rho}\Big)\nonumber\\
   & +\frac{4r_G^4}{9m_b^4}\Big(16-21\rho+9\rho^2-7\rho^3+3\rho^4+12\log{\rho}\Big)
    +\frac{2\tilde{s}_E^4}{9m_b^4}\Big(25-36\rho+20\rho^3-9\rho^4+12\log{\rho}\Big)\nonumber\\
   & +\frac{2s_B^4}{3m_b^4}\Big((\rho-1)^3(5\rho+1)\Big)
    -\frac{s_{qB}^4}{36m_b^4}\Big(25-48\rho+36\rho^2-16\rho^3+3\rho^4+12\log{\rho}\Big)\nonumber\\
   & -\frac{4X_1^5}{15m_b^5}\Big((\rho -1)^2 \left(72 \rho ^2+29 \rho +7\right)\Big)  \nonumber\\
  & +\frac{X_2^5}{90m_b^5}\Big(279 \rho ^4-400 \rho ^3+180 \rho ^2-420 \log{\rho}+85-\frac{144}{\rho }\Big)\nonumber\\
  &  -\frac{X_3^5}{5m_b^5}\Big(-9 \rho ^4+30 \rho ^3-20 \rho ^2-20 \rho -20 \log{\rho}+45-\frac{26}{\rho }\Big)\nonumber\\
   & +\frac{X_4^5}{90m_b^5}\Big(-63 \rho ^4+200 \rho ^3-180 \rho ^2-60 \log{\rho}+115-\frac{72}{\rho }\Big)\nonumber\\
  &  +\frac{4X_5^5}{m_b^5}\Big( (\rho -1)^2 \left(1-\rho-2\rho^2\right)\Big)\nonumber\\
  &   -\frac{4X_6^5}{9m_b^5}\Big(-18 \rho ^4+34 \rho ^3-9 \rho ^2-18 \rho +6 \log{\rho}+11\Big)\nonumber\\
   & +\frac{2X_7^5}{9m_b^5}\Big( \rho ^3+6 \rho ^2-36 \rho +24 \log{\rho}+26+\frac{3}{\rho }\Big)\nonumber\\
   & +\frac{4X_8^5}{3m_b^5}\Big(-3 \rho ^4+8 \rho ^3-5 \rho ^2-3 \rho +4-\frac{1}{\rho }\Big)\nonumber\\
  &  -\frac{X_9^5}{9m_b^5}\Big(27 \rho ^4-61 \rho ^3+30 \rho ^2+12 \log{\rho}+7-\frac{3}{\rho }\Big)\nonumber\\
  &  +\frac{X_{10}^5}{9m_b^5}\Big(9 \rho ^4-19 \rho ^3+6 \rho ^2+12 \log{\rho}+1+\frac{3}{\rho }\Big) +\mathcal{O}\Big(\frac{1}{m_b^6}\Big)\ .
\end{align}
\clearpage

\section{\boldmath Centralised $q^2$ moments \unboldmath}
\label{sec:ap_q2moments}
In this appendix, we present numerical values for the centralized $q^2$ moments. The full expressions including a $\hat{q}^2_{\rm cut}$ can be obtained from the Mathematica notebook added as an ancillary file (see Appendix \ref{sec:ap_traceform} for more details). We employ the following values \cite{Bordone:2021oof}:
\begin{align}
    q^2_{\rm{cut}}=0\ \text{GeV}^2\ , \hspace{1cm}m_b^{\rm{kin}}=4.573\ \text{GeV}\ ,\hspace{1cm}\bar{m}_c(2\ \text{GeV})=1.092\ \text{GeV}\ .
\end{align} 
We then find\footnote{This corrects for a typo in the coefficient of $s_{qB}^4$ in $q_3$ in \cite{Fael:2018vsp}. The other rounding differences arise due to the higher precision used here for the quark masses.}
\begin{align}
q_1=&\ \frac{m_b^2}{\mu_3}\Big(0.22\mu_3-0.57\frac{\mu_G^2}{m_b^2}-1.4\frac{(\mu_G^2)^2}{m_b^4\mu_3}-5.5\frac{\tilde{\rho}_D^3}{m_b^3}+16\frac{\tilde{r}_E^4}{m_b^4}-5.7\frac{r_G^4}{m_b^4}-1.7\frac{\tilde{s}_E^4}{m_b^4}\nonumber\\&+0.097\frac{s_B^4}{m_b^4}-0.064\frac{s_{qB}^4}{m_b^4}-24\frac{\mu_G^2\tilde{\rho}_D^3}{m_b^5\mu_3}-19\frac{X_1^5}{m_b^5}+18\frac{X_2^5}{m_b^5}-15\frac{X_3^5}{m_b^5}+2.3\frac{X_4^5}{m_b^5}\nonumber\\&+6.5\frac{X_5^5}{m_b^5}+0.91\frac{X_6^5}{m_b^5}-7.0\frac{X_7^5}{m_b^5}+3.1\frac{X_8^5}{m_b^5}+5.2\frac{X_9^5}{m_b^5}-2.0\frac{X_{10}^5}{m_b^5}+0.047\frac{X_{\rm{IC}}^{5}}{m_b^3m_c^2}\Big)\ ,\nonumber\\
q_2=&\ \frac{m_b^4}{\mu_3}\Big(0.022\mu_3-0.12\frac{\mu_G^2}{m_b^2}-0.61\frac{(\mu_G^2)^2}{m_b^4\mu_3}-1.6\frac{\tilde{\rho}_D^3}{m_b^3}+7.7\frac{\tilde{r}_E^4}{m_b^4}-2.1\frac{r_G^4}{m_b^4}-0.66\frac{\tilde{s}_E^4}{m_b^4}\nonumber\\&+0.20\frac{s_B^4}{m_b^4}-0.082\frac{s_{qB}^4}{m_b^4}-12\frac{\mu_G^2\tilde{\rho}_D^3}{m_b^5\mu_3}-20\frac{X_1^5}{m_b^5}+15\frac{X_2^5}{m_b^5}-22\frac{X_3^5}{m_b^5}+3.2\frac{X_4^5}{m_b^5}\nonumber\\&+4.2\frac{X_5^5}{m_b^5}-0.32\frac{X_6^5}{m_b^5}-4.9\frac{X_7^5}{m_b^5}+5.4\frac{X_8^5}{m_b^5}+1.8\frac{X_9^5}{m_b^5}-1.2\frac{X_{10}^5}{m_b^5}+0.030\frac{X_{\rm{IC}}^{5}}{m_b^3m_c^2}\Big)\ ,\nonumber\\
q_3=&\ \frac{m_b^6}{\mu_3}\Big(0.0012\mu_3-0.013\frac{\mu_G^2}{m_b^2}-0.24\frac{(\mu_G^2)^2}{m_b^4\mu_3}-0.34\frac{\tilde{\rho}_D^3}{m_b^3}+2.9\frac{\tilde{r}_E^4}{m_b^4}-0.56\frac{r_G^4}{m_b^4}-0.19\frac{\tilde{s}_E^4}{m_b^4}\nonumber\\&+0.093\frac{s_B^4}{m_b^4}-0.035\frac{s_{qB}^4}{m_b^4}-5.8\frac{\mu_G^2\tilde{\rho}_D^3}{m_b^5\mu_3}-12\frac{X_1^5}{m_b^5}+9.3\frac{X_2^5}{m_b^5}-17\frac{X_3^5}{m_b^5}+2.5\frac{X_4^5}{m_b^5}\nonumber\\&+2.0\frac{X_5^5}{m_b^5}-0.42\frac{X_6^5}{m_b^5}-2.8\frac{X_7^5}{m_b^5}+4.4\frac{X_8^5}{m_b^5}+0.10\frac{X_9^5}{m_b^5}-0.70\frac{X_{10}^5}{m_b^5}+0.016\frac{X_{\rm{IC}}^{5}}{m_b^3m_c^2}\Big)\ ,\nonumber\\
q_4=&\ \frac{m_b^8}{\mu_3}\Big(0.0010\mu_3-0.012\frac{\mu_G^2}{m_b^2}-0.10\frac{(\mu_G^2)^2}{m_b^4\mu_3}-0.22\frac{\tilde{\rho}_D^3}{m_b^3}+1.6\frac{\tilde{r}_E^4}{m_b^4}-0.33\frac{r_G^4}{m_b^4}-0.11\frac{\tilde{s}_E^4}{m_b^4}\nonumber\\&+0.047\frac{s_B^4}{m_b^4}-0.018\frac{s_{qB}^4}{m_b^4}-2.6\frac{\mu_G^2\tilde{\rho}_D^3}{m_b^5\mu_3}-7.8\frac{X_1^5}{m_b^5}+7.6\frac{X_2^5}{m_b^5}-17\frac{X_3^5}{m_b^5}+2.5\frac{X_4^5}{m_b^5}\nonumber\\&+1.2\frac{X_5^5}{m_b^5}-0.26\frac{X_6^5}{m_b^5}-2.5\frac{X_7^5}{m_b^5}+4.3\frac{X_8^5}{m_b^5}-0.40\frac{X_9^5}{m_b^5}-0.84\frac{X_{10}^5}{m_b^5}+0.015\frac{X_{\rm{IC}}^{5}}{m_b^3m_c^2}\Big)\ .\label{numeric_qi}
\end{align}
We stress that even though $X_{\rm{IC}}^5$ has a small prefactor, it consists of a linear combination of the $X_i^5$ HQE parameters, and thus the value of $X_{\rm{IC}}^5$ itself may be much larger compared to the individual $X_i^5$ and therefore the IC contribution to the $q^2$ moments may become significant despite the small prefactor. This depends on the signs of the $X_i^5$, but in the LLSA we find that $X_{\rm{IC}}^5\approx 14.71$ GeV$^5$. Finally, one can write the expression for the $q^2$ moments using only $X_{1...10}^5$ through the following replacement:
\begin{align}
    \frac{X^{5}_{\rm{IC}}}{m_c^2}&\mapsto\Big(\frac{4.573}{1.092}\Big)^2\frac{1}{m_b^2}\Big(-24X_2^5+78X_3^5-12X_4^5+10X_7^5-20X_8^5+5X_9^5+5X_{10}^5\Big)\ .
\end{align}

\section{LLSA expressions}\label{sec:ap_LLSA}
Using the LLSA for the ``perp"-basis matrix elements from \cite{Heinonen:2014dxa}, we can find LLSA expressions for our RPI-basis matrix elements. These expressions are functions of $\epsilon_{1/2}$, $\epsilon_{3/2}$, and the ``perp"-matrix elements \cite{Heinonen:2014dxa} 
\begin{align}
    2m_B\muPperp&=-\langle \bar{b}_v\, (iD_\mu)\, (iD_\nu)\, b_v\rangle g^{\mu\nu}_\perp\nonumber\ ,\\
    2m_B\muGperp&=\langle \bar{b}_v\, (iD_\alpha)\, (iD_\beta)\, (-i\sigma_{\mu\nu})\,b_v\rangle g_\perp^{\mu\alpha}g_\perp^{\nu\beta}\ ,
\end{align}
where $g^\perp_{\mu\nu}\equiv g_{\mu\nu}-v_\mu v_\nu$ and where we have dropped the usual power of 2 in the definitions of $\big(\mu_{\pi(G)}^2\big)^\perp$ to simplify the expressions in this appendix. The LLSA expressions up to $1/m_b^5$ we find for the RPI operators and IC operator are given by
\begin{align}
    \mu_3=&\ 1+\frac{1}{2m_b^2}\Big[\muGperp-\muPperp\Big]+\frac{\epsilon_{1/2}}{2m_b^3}\Big[\muGperp-\muPperp\Big]+\frac{1}{8m_b^4}\Big[(\muGperp-\muPperp)(\muPperp-\muGperp+3\epsilon_{1/2}^2)\Big]\nonumber\\
   &+\frac{\epsilon_{1/2}}{8m_b^5}\Big[(\muGperp-\muPperp)(\muPperp-\muGperp+4\epsilon_{1/2}^2)\Big]\ ,\nonumber\\
   \mu_G^2=&\ \muGperp+\frac{\epsilon_{1/2}}{m_b}\Big[\muGperp-\muPperp\Big]+\frac{1}{2m_b^2}\Big[(\muGperp-\muPperp)(\muPperp-\muGperp+\epsilon_{1/2}^2)\Big]\nonumber\\
   &+\frac{\epsilon_{1/2}}{2m_b^3}\Big[(\muGperp-\muPperp)(\muPperp-\muGperp+2\epsilon_{1/2}^2)\Big]\ ,\nonumber\\
   \tilde{\rho}_D^3=&\ -\frac{1}{3}\Big[\epsilon_{1/2}(\muGperp-\muPperp)-\epsilon_{3/2}(\muGperp+2\muPperp)\Big]\nonumber\\
   &+\frac{1}{12m_b}\Big[2\epsilon_{1/2}^2(\muGperp-\muPperp)-\epsilon_{3/2}^2(\muGperp+\muPperp)+(\muGperp)^2-6\muGperp\muPperp+2(\muPperp)^2\Big]\nonumber\\
   &+\frac{1}{12m_b^2}\Big[-\epsilon_{1/2}(\muGperp-\muPperp)(4\muPperp-\muGperp+3\epsilon_{1/2}^2)\Big]\ ,\nonumber
 \\
   \tilde{r}_E^4=&\ -\frac{1}{3}\Big[\epsilon_{1/2}^2(\muGperp-\muPperp)-\epsilon_{3/2}^2(\muGperp+2\muPperp)\Big]\nonumber\\
   &+\frac{1}{18m_b}\Big[2\epsilon_{1/2}(4(\muGperp)^2-5\muGperp\muPperp+(\muPperp)^2)-\epsilon_{3/2}(5(\muGperp)^2+8\muGperp\muPperp-4(\muPperp)^2)\Big]\ ,\nonumber  \end{align}
   
   \begin{align}
   r_G^4=&\ -\frac{2}{3}\Big[(\muGperp)^2+\epsilon_{1/2}^2(\muGperp-\muPperp)-\epsilon_{3/2}^2(\muGperp+2\muPperp)\Big]\nonumber\\
   &+\frac{2}{3m_b}\Big[(\muGperp-\muPperp)(\epsilon_{1/2}(\muGperp-\muPperp)-\epsilon_{3/2}(\muGperp+2\muPperp))\Big]\ ,\nonumber\\
   \tilde{s}_E^4=&\ -\frac{1}{3}\Big[2\epsilon_{1/2}^2(\muGperp-\muPperp)+\epsilon_{3/2}^2(\muGperp+2\muPperp)\Big] \nonumber\\
   &+\frac{1}{18m_b}\Big[4\epsilon_{1/2}(4(\muGperp)^2-5\muGperp\muPperp+(\muPperp)^2)\nonumber\\
   &+\epsilon_{3/2}(5(\muGperp)^2+8\muGperp\muPperp-4(\muPperp)^2)-18\epsilon^3_{1/2}(\muGperp-\muPperp)\Big]\ ,\nonumber\\
   s_B^4=&\ -\frac{1}{3}\Big[2\epsilon_{1/2}^2(\muGperp-\muPperp)+\epsilon_{3/2}^2(\muGperp+2\muPperp)+2(\muGperp)^2\Big]\nonumber\\
   &-\frac{1}{9m_b}\Big[(\muGperp-\muPperp)(2\epsilon_{1/2}(2\muGperp+\muPperp)-4\epsilon_{3/2}(\muGperp+2\muPperp)+9\epsilon_{1/2}^3)\Big]\ ,\nonumber\\
   s_{qB}^4=&\ -\frac{2}{3}\muGperp(\muGperp+10\muPperp)\nonumber\\
   &-\frac{2}{9m_b}\Big[(\muGperp-\muPperp)(9\epsilon_{1/2}^3-\epsilon_{1/2}(\muGperp-10\muPperp)+4\epsilon_{3/2}(\muGperp+2\muPperp))\Big]\ ,\nonumber\\
    X_1^5=&\ -\frac{1}{3}\Big[\epsilon_{1/2}^3(\muGperp-\muPperp)-\epsilon_{3/2}^3(\muGperp+2\muPperp)\Big]\ ,\nonumber\\
    X_2^5=&\ 0\ ,\nonumber\\
    X_3^5=&\ \frac{1}{18}\Big[2\epsilon_{1/2}(3(\muGperp)^2-5\muGperp\muPperp+2(\muPperp)^2)+\epsilon_{3/2}(3(\muGperp)^2+10\muGperp\muPperp+8(\muPperp)^2)\Big]\ ,\nonumber\\
    X_4^5=&\ \frac{1}{9}\Big[2\epsilon_{1/2}((\muGperp)^2+5\muGperp\muPperp-6(\muPperp)^2)+\epsilon_{3/2}((\muGperp)^2-10\muGperp\muPperp-24(\muPperp)^2)\Big]\ ,\nonumber\\
    X_5^5=&\ -\frac{1}{3}\Big[2\epsilon_{1/2}^3(\muGperp-\muPperp)+\epsilon_{3/2}^3(\muGperp+2\muPperp)\Big]\ ,\nonumber\\
    X_6^5=&\ 0\ ,\nonumber
  \\
    X_7^5=&\ -\frac{4}{9}\muGperp\Big[\epsilon_{1/2}(\muGperp-\muPperp)-\epsilon_{3/2}(\muGperp+2\muPperp)\Big]\ ,\nonumber\\
    X_8^5=&\ \frac{1}{18}\Big[4\epsilon_{1/2}((\muGperp)^2-3\muGperp\muPperp+2(\muPperp)^2)+\epsilon_{3/2}(5(\muGperp)^2+6\muGperp\muPperp-8(\muPperp)^2)\Big]\ ,\nonumber  \\
    X_9^5=&\ -\frac{4}{3}\muGperp\Big[\epsilon_{1/2}(\muGperp-\muPperp)-\epsilon_{3/2}(\muGperp+2\muPperp)\Big]\ ,\nonumber\\
    X_{10}^5=&\ -\frac{1}{9}\Big[\epsilon_{1/2}(6(\muGperp))^2-2\muGperp\muPperp-4(\muPperp)^2)-\epsilon_{3/2}(3(\muGperp)^2+4\muGperp\muPperp-4(\muPperp)^2)\Big]\ ,\nonumber\\
    X_{\rm{IC}}^{5}=&\ \frac{10}{9}\Big[4\epsilon_{1/2}((\muGperp)^2-7\muGperp\muPperp+6(\muPperp)^2)+\epsilon_{3/2}(17(\muGperp)^2+67\muGperp\muPperp+66(\muPperp)^2)\Big]\ .
\end{align}

\bibliographystyle{jhep} 
\bibliography{refs.bib} 

\end{document}